\documentclass[useAMS,usenatbib]{mn2e}
\usepackage{graphicx}
\usepackage{comment}
\usepackage{color} 

\title[Structure for quasars and polarisation I. The polarisation dichotomy]
      {A structure for quasars under the scope of polarisation \\ 
       I. The UV/optical polarisation dichotomy of type-1 and type-2 AGN}

\author[F. Marin and R. W. Goosmann]{F. Marin\thanks{E-mail: frederic.marin@astro.unistra.fr} and R. W. Goosmann\\
Observatoire Astronomique de Strasbourg, Universit\'e de Strasbourg, CNRS, UMR 7550, 11 rue de l'Universit\'e, 67000 Strasbourg, France}

\begin{document}

\date{Accepted 2013 September 14. Received 2013 September 13; in original form 2013 June 30}
\pagerange{\pageref{firstpage}--\pageref{lastpage}} \pubyear{2013}

\maketitle

\begin{abstract}
We present UV/optical spectropolarimetric modelling of the phenomenologically-based structure for quasars proposed by \citet{Elvis2000}. In this first 
paper of a series, we explore the continuum polarisation emerging from radiatively accelerated and bent winds that were vertically launched from 
the accretion disc in an active galactic nucleus (AGN). We simulate the radiative transfer occurring in Thomson scattering and dust extinction media 
over a range of morphological parameters and optical depths of the wind. We demonstrate that the wind geometry proposed by Elvis with a phenomenologically-derived 
bending angle of $\theta = 60^\circ$ still underestimates the observed optical polarisation percentage of type-1 and type-2 AGN and does not yet reproduce the 
expected dichotomy of the polarisation position angle. To recover the observed polarisation properties, a smaller bending angle and some amount of dust 
shielding in the equatorial region should be considered. A two-phase outflow is found to generate both the observed polarisation dichotomy and acceptable 
levels of polarisation degree if the wind has a bending angle $\theta$ = 45$^\circ$, and the conical shells have a half-opening angle of 
3$^\circ < \delta\theta < $ 10$^\circ$. The absorbing dust column at the wind base should be in the range of $1 < \tau_{dust} \le 4$ ($\tau$ being integrated 
over 2000 -- 8000~\AA). Straightforward observational tests from spectropolarimetry and from determining the number density of different AGN types can be performed 
to further constrain the wind geometry.

\end{abstract}


\label{firstpage}

\section{Introduction}
\label{Sect1}

While the topic of quasar research started in the beginning of the twentieth century \citep{Fath1909}, it took nearly ninety years to built up a 
coherent, unifying model for radio-quiet AGN\footnote{In this paper we only consider radio-quiet AGN models, as the presence of a relativistic 
jet would lead to a more complicated spectropolarimetric picture by adding intrinsically highly polarised synchrotron emission} \citep{Antonucci1993}. 
It is now widely accepted that a supermassive black hole (SMBH) with mass ranging from 10$^6$ to 10$^{10}$ M$_\odot$ is surrounded by an accretion 
disc and its associated corona, who radiate a continuum spectra from the soft X-ray energy to the near-infrared waveband. Broad line signatures are 
thought to arise from high velocity gas trapped close to the equatorial plane between the accretion disc and a surrounding, optically thick, circumnuclear 
dust medium. This "dusty torus" is thought to collimate ejection winds along the polar direction, where narrow line features are expected to be 
produced by distant, low velocity outflows. 

According to the unified scheme, many of the AGN observational characteristics can be explained by an inclination effect rather than intrinsic 
composition differences. However, the exact morphology of the inner AGN regions remains highly debated due to a wide panel of observational 
emission and absorption line features. If cylinder-like, spinning, equatorial reprocessing regions where initially used to model the broadening
of the emission lines in Seyfert-1 objects (see \citealt{Osterbrock1991} for a review), it is now admitted that the morphology of the broad line region 
is by far more complex (e.g. \citealt{Davidson1979,Mathews1985,Shields1977,Eracleous2006,Gaskell2009}). In this picture, the central, ionising source 
is surrounded by a roughly spherical distribution of clumps in Keplerian motion, situated at the outermost areas of the accretion disc \citep{Gaskell2013}.
Similar assumptions about the clumpy nature of the narrow line regions \citep{Capetti1995,Ogle2003} increase the difficulty of the challenging but necessary 
goal of drawing a unifying model for the sub-parsec regions of quasars.

In this context, \citet{Elvis2000} developed a phenomenologically-based model to explain the wide variety of emission and absorption features in AGN spectra, 
assuming a simple outflowing structure. The model presented by \citet{Elvis2000} assumes a flow of warm, highly ionised matter (WHIM) that is launched from 
an accretion disc over a small range of radii and then bent outward and driven into a radial direction by radiation pressure (see Fig. 1 and \citealt{Elvis2000}). 
Certain aspects of this phenomenologically-derived structure can be directly tested from number counts of type-1 and type-2 AGN or the relative number of broad 
absorption line (BAL) versus narrow absorption line (NAL) objects. The wind structure surrounds the irradiating accretion disc and gives rise to a reprocessing 
spectrum that varies with the disc luminosity and the viewing angle of the observer. Time-resolved spectroscopy therefore is a powerful tool to test the model 
against the observations, but this technique may not be sufficiently sensitive to the outflow geometry and dynamics. 

To extend the comparison of the outflow model to the observations somewhat further, spectropolarimetric data of AGN can be explored. In the optical/UV band, 
the polarisation of radio-quiet AGN is mostly determined by reprocessing and strongly depends on the scattering geometry. We expect BAL objects to be 
generally more highly polarized than non-BAL QSOs \citep{Ogle1999} and NAL Seyfert-like galaxies to show polarisation parallel to the axis of the torus
associated with polarisation degrees inferior to unity \citep{Smith2002}. Spectropolarimetric observations of nearby Seyfert galaxies showed that there 
is an observational dichotomy between type-1 and type-2 AGN: a large fraction of AGN seen by the pole (type-1) show a polarisation position angle oriented 
parallel to the system axis, while edge-on objects (type-2) exhibit a perpendicular polarisation position angle \citep{Antonucci1983}. When a polarisation 
model is provided, one of its goals is to reproduce the observed dichotomy, where parallel polarisation is thought to originate close to the 
SMBH, while perpendicular polarisation emerges from the circumnuclear matter or the ionised winds (see references and modelling in \citealt{Goosmann2007,Marin2012a}). 
Direct \textit{Hubble Space Telescope} (HST) polarization imaging of NGC~1068 and Markarian~477 showed that perpendicular polarisation can also emerge from scattering 
clumps situated 10 to 100 pc away from the innermost regions \citep{Capetti1995,Kishimoto1999,Kishimoto2002}; in these cases, the necessity to reproduce type-2 
perpendicular polarisation is less critical. However, the production of parallel polarisation in type-1 AGN is a rather important challenge, as it might give
us an insight of the intrinsic spectral shape of the central engine. The detection of a Balmer absorption edge in a few quasars \citep{Kishimoto2003,Kishimoto2004}, 
thought to originates interior to the BLR, strengthens the role of parallel polarisation as a tool to constrain the geometry and the composition of reprocessing and 
scattering media close to the SMBH \citep{Kishimoto2008}. This continues to be very relevant in AGN research where geometry is a key parameter for unification theories 
\citep{Antonucci1985, Antonucci1993, Elvis2000}.

In the present work, we model the polarisation properties produced by radiative reprocessing inside the structure of quasars as suggested by 
\citet{Elvis2000}. We focus on the continuum polarisation at optical and ultraviolet wavelengths and we test if the outflow model can reproduce the 
observed dichotomy with respect to the polarisation position angle \citep{Antonucci1983,Antonucci1984,Smith2002}. The remainder of the paper is organised 
as follows: in Sect.~\ref{Sect2} we present and analyse spectropolarimetric simulations of disc-born outflows for different model parameters. 
In Sect.~\ref{Sect3} we discuss our results and relate them to the observational constraints before drawing our conclusions in Sect.~\ref{Sect4}.


\section{Exploring the structure for quasars by spectropolarimetry}
\label{Sect2}

\subsection{Radiative transfer code and model geometry}

\begin{figure*}
\centering
   \includegraphics[trim = 0mm 145mm 5mm 40mm, clip, width=18cm]{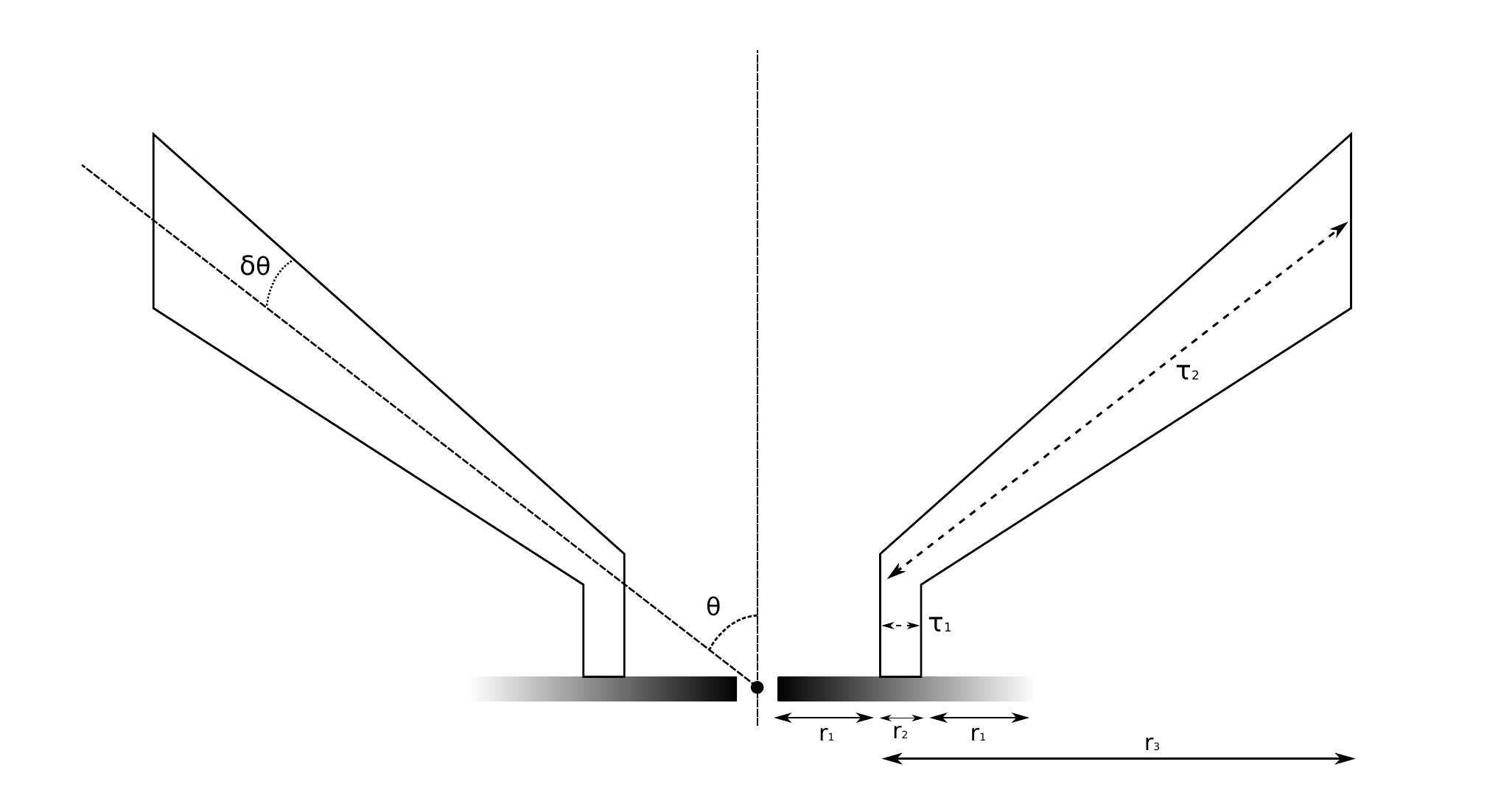}
   \caption{Schematic view of the structure proposed by \citet{Elvis2000} and
            implemented in the {\sc stokes} model.
            The outflow arises vertically from the accreting disc and is 
            bent outward by radiation pressure along a 60$^\circ$ direction
            relative to the model symmetry axis.
            The half-opening angle of the wind extension is 3$^\circ$. The radial 
            optical depths of the wind base and of the outflowing material
            are set to by $\tau_1$ and $\tau_2$ respectively.}
  \label{Fig1.1}
\end{figure*}

To study the model of \citet{Elvis2000}, we apply the version 1.2 of Monte Carlo radiative transfer code {\sc stokes} presented in \citet{Goosmann2007} and upgraded 
by \citet{Marin2012a}. This modelling suite coherently treats three-dimensional transfer and multiple reprocessing between emitting and scattering regions 
and includes a treatment of polarisation. The model space is surrounded by a spherical web of virtual detectors. The detectors record the wavelength, 
intensity and polarisation state of each photon. The latest-to-date version of {\sc stokes} also generates polarisation images with the photons being 
projected onto the observer's plane of the sky and then stored in planar coordinates. The net intensity, polarisation degree $P$ and polarisation position 
angle $\psi$ as a function of wavelength are then computed by summing up the Stokes vectors of all detected photons in a given spectral and spatial bin. 
The spectra can be evaluated at each viewing direction in the polar and azimuthal direction $i$ and $\phi$. Note that a position angle $\psi = 0\degr$ denotes 
a polarisation state with the $\vec E$-vector oscillating in a perpendicular direction with respect to the symmetry axis of the wind, while for $\psi = 90\degr$ 
the $\vec E$-vector is parallel to the (projected) wind axis. 

We simulate the radiative transfer through the structure for quasars as proposed by \citet{Elvis2000} by constructing continuous outflow regions that can be filled with electrons 
or dust or a combination of the two at uniform density. Assuming a uniform density certainly is an oversimplification and we are going to release this 
condition in future work. For now, we only perform the most basic polarisation test of the wind model. For all models presented in this paper, the equatorial 
emitting region will be defined as an isotropic, disc-like source emitting an unpolarised spectrum with a power-law spectral energy distribution 
$F_{\rm *}~\propto~\nu^{-\alpha}$ and $\alpha = 1$. The model geometry is summarised in Fig.~\ref{Fig1.1} : the wind arises from the accretion disc at a 
distance r$_1$ of 0.0032~pc (10$^{16}$ cm) and is bent into a direction of $\theta$ = 60$^\circ$ relative to the model symmetry axis. The range of radii 
where the flow arises from the accretion disc is parametrised by r$_2$ = 0.00032~pc (10$^{15}$ cm). Finally, the half-opening angle $\delta\theta$ of the 
wind is 3$^\circ$ and the flow extends up to r$_3$ = 0.032~pc (10$^{17}$ cm). It is instructive to express these distances in units of the gravitational 
radius ($R_{\rm G}$) for a fiducial black hole mass of $\sim$ 10$^7$~M$_\odot$ (such as in NGC~5548), as well as in units of the dust sublimation radius 
($T_{\tau, \rm K}$, $T$ = 1500~K, \citealt{Suganuma2006,Kishimoto2007}) and BLR radius ($R_{\rm BLR}$, \citealt{Bentz2009}). We estimated $T_{\tau, \rm K}$ 
and $R_{\rm BLR}$ from NGC~5548's near-IR reverberation measurements and reverberation mapping of H$\beta$, respectively. Converted radii, given in Tab.~\ref{Tab1},
imply that the phenomenological outflow derived by \citet{Elvis2000} arises below the dust formation radius and the BLR is likely to form closely beyond the 
inner part of the wind. The extension of the wind covers a large fraction of the BLR, essentially obscuring the highly ionized clouds close to the central engine.

\begin{table}
  \centering
  {
    \begin{tabular}{l*{4}{l}l}
    Radius      & pc 	  & cm 	      & $R_{\rm G}$ & $T_{\tau, \rm K}$ & $R_{\rm BLR}$ \\
    \hline
    r$_1$ 	& 0.0032  & 10$^{16}$ & 6684 	    & 0.08 		& 0.91  \\
    r$_2$   	& 0.00032 & 10$^{15}$ & 668 	    & 0.008 		& 0.091  \\
    r$_3$   	& 0.032   & 10$^{17}$ & 66840	    & 0.8		& 9.1    \\
    \end{tabular}
  }
  \caption{Parameters of the model expressed in units of gravitational radius $R_{\rm G}$
	  (for a black hole mass of $\sim$ 10$^7$~M$_\odot$), dust sublimation radius $T_{\tau, \rm K}$ and
	  BLR radius $R_{\rm BLR}$. See text for further details.}
  \label{Tab1}
\end{table}

\subsection{The warm, highly ionised medium}
\label{Sect2.1}

\subsubsection{Testing the electron-dominated outflows}
\label{Sect2.1.1}

We assume that the WHIM is mainly composed of ionised gas with a Thomson optical depth $\tau$ of the order of unity along the conical outflowing direction. 
This wind extends out to several parsec and can be associated with the "ionisation cones" observed in nearby Seyfert-2 galaxies \citep{Capetti1995,Axon1996,Osterbrock1991}. 
To investigate the scattering properties and the resulting spectra and polarisation from the WHIM, we first consider a uniform-density, continuous medium made 
of electrons. The Thomson optical depth at the wind base and inside the conical, outflowing direction are set to $\tau_1 \sim$~0.02 ($n_{\rm e}$ = 3$\times$10$^7$)
and $\tau_2 \sim$~2 ($n_{\rm e}$ = 2.6$\times$10$^6$), respectively. Such an optical depth is in agreement with observational measurements \citep{Packham1997,Ogle2003} 
and also suggested by polarisation modelling work on NGC~1068 \citep{Miller1991,Young1995,Goosmann2011}.

\begin{figure}
\centering
   \includegraphics[trim = 10mm 0mm 0mm 0mm, clip, width=10cm]{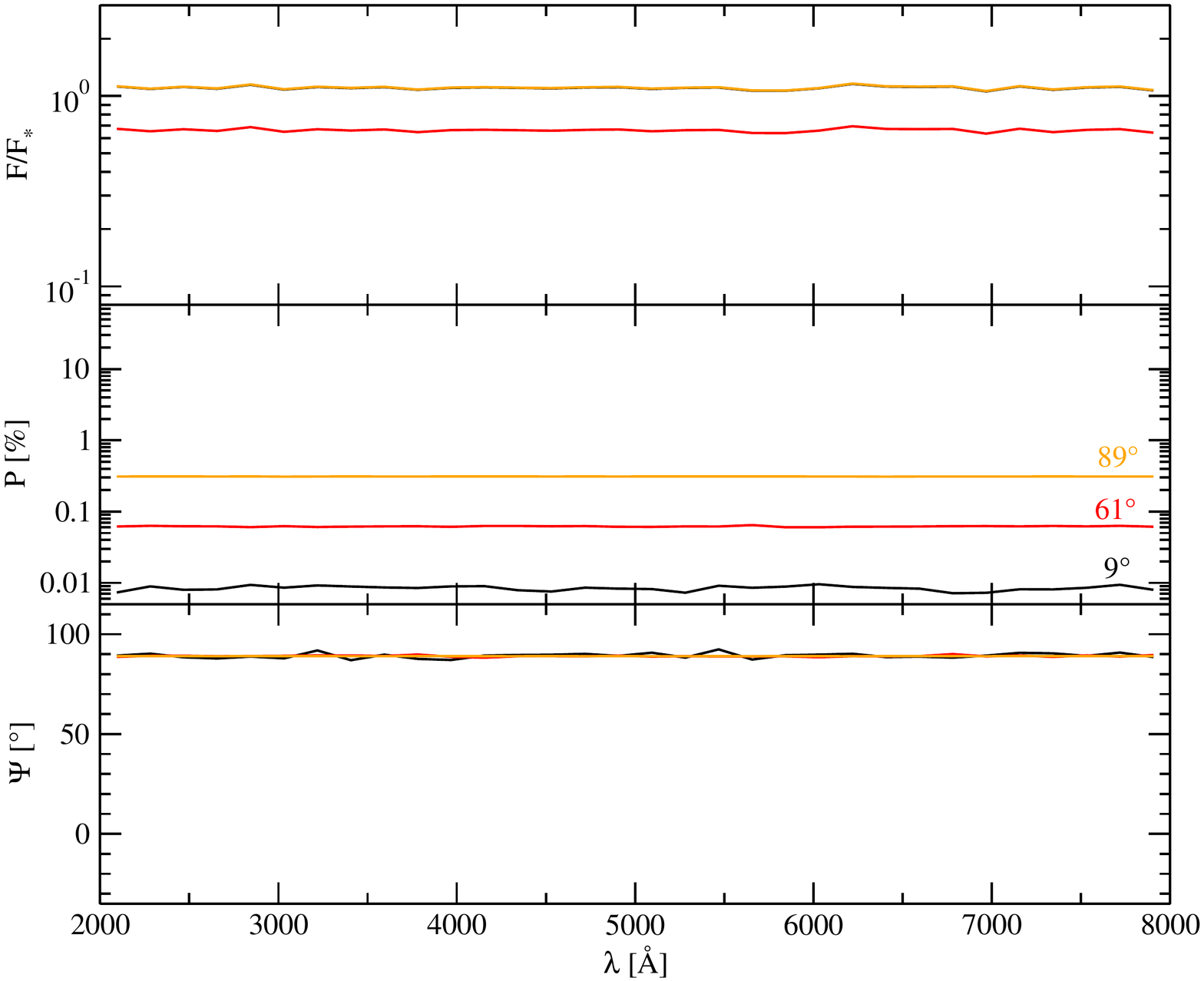}
   \caption{Modelling an electron-filled structure for quasars outflow
            using an uniform model such as in Fig.~\ref{Fig1.1},
            seen at different viewing inclinations, \textit{i} 
            (black : 9$^\circ$ -- red : 61$^\circ$ -- orange : 89$^\circ$).
            \textit{Top}: the fraction, $F$/$F_{\rm *}$ of the central flux;
            \textit{middle}: polarisation, \textit{P};
            \textit{bottom}: polarisation angle, $\psi$.}
  \label{Fig1.2}
\end{figure}

In Fig.~\ref{Fig1.2}, we plot the simulated spectropolarimetric results across the 2000 -- 8000~\AA~wave band and as a function of the observer's viewing angle. 
We consider three different lines-of-sight : one along a polar inclination (9$^\circ$), one passing through the outflowing wind (61$^\circ$), 
and one along an extreme equatorial inclination (89$^\circ$). The inclination of the system is defined with respect to the symmetry axis of the model. 
The polar and equatorial lines of sight do not cross the extended wind and the fraction $F/F_{\rm *}$ of the central flux, $F_{\rm *}$, remains similar for both 
inclinations. At intermediate inclinations, however, a significant fraction of the radiation is scattered out of the line-of-sight by the optically thick wind. 
The polarisation percentage at all three viewing directions is wavelength-independent, as it is expected for Thomson scattering, and the polarisation position 
angle is $\psi = 90\degr$ (parallel polarisation) for all cases. The polarisation percentage $P$ is very low at polar viewing angles and then rises with $i$. 
It increases to 0.05~\% when looking through the extended wind and then further to about 0.3~\% for an edge-on line-of-sight.

\begin{figure}
\centering
   \includegraphics[trim = 0mm 5mm 0mm 10mm, clip, width=8cm]{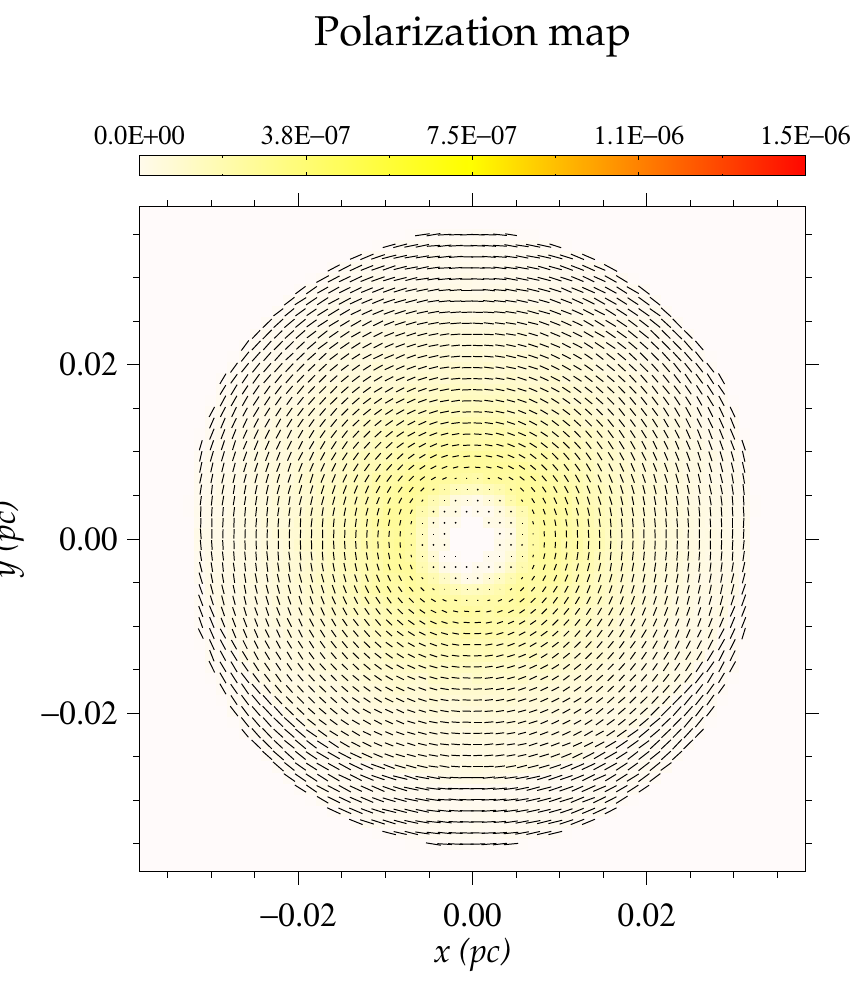}
   \includegraphics[trim = 0mm 5mm 0mm 18mm, clip, width=8cm]{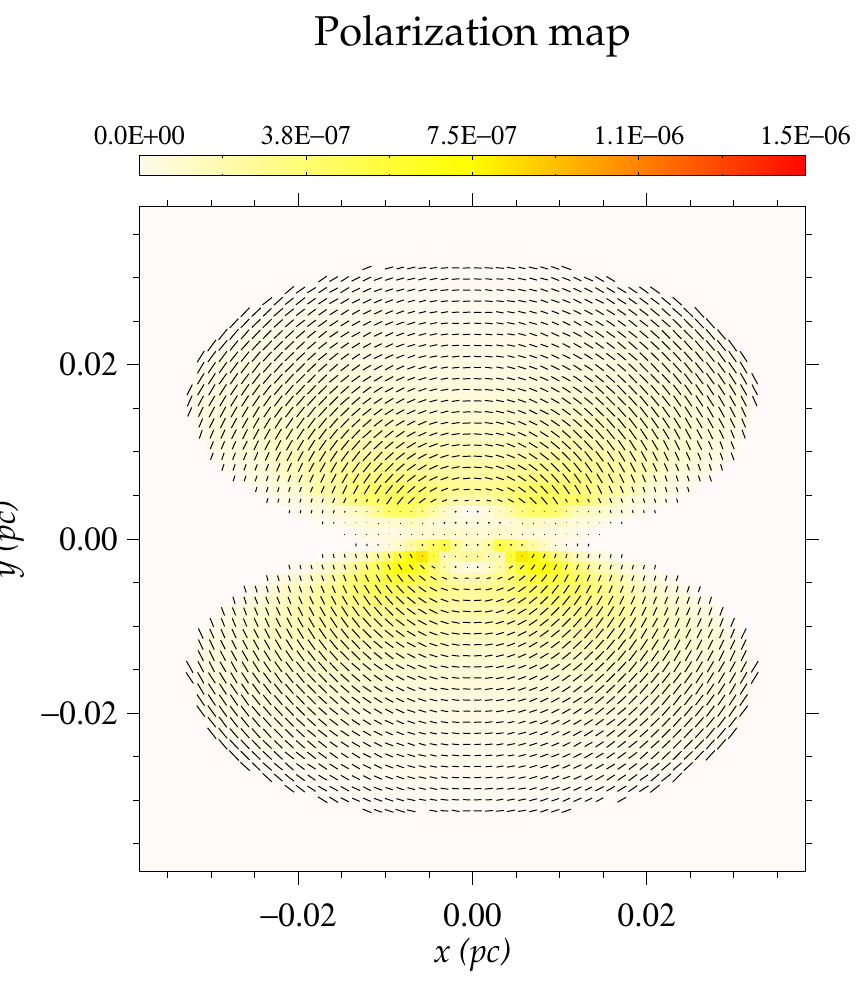}
   \includegraphics[trim = 0mm 0mm 0mm 18mm, clip, width=8cm]{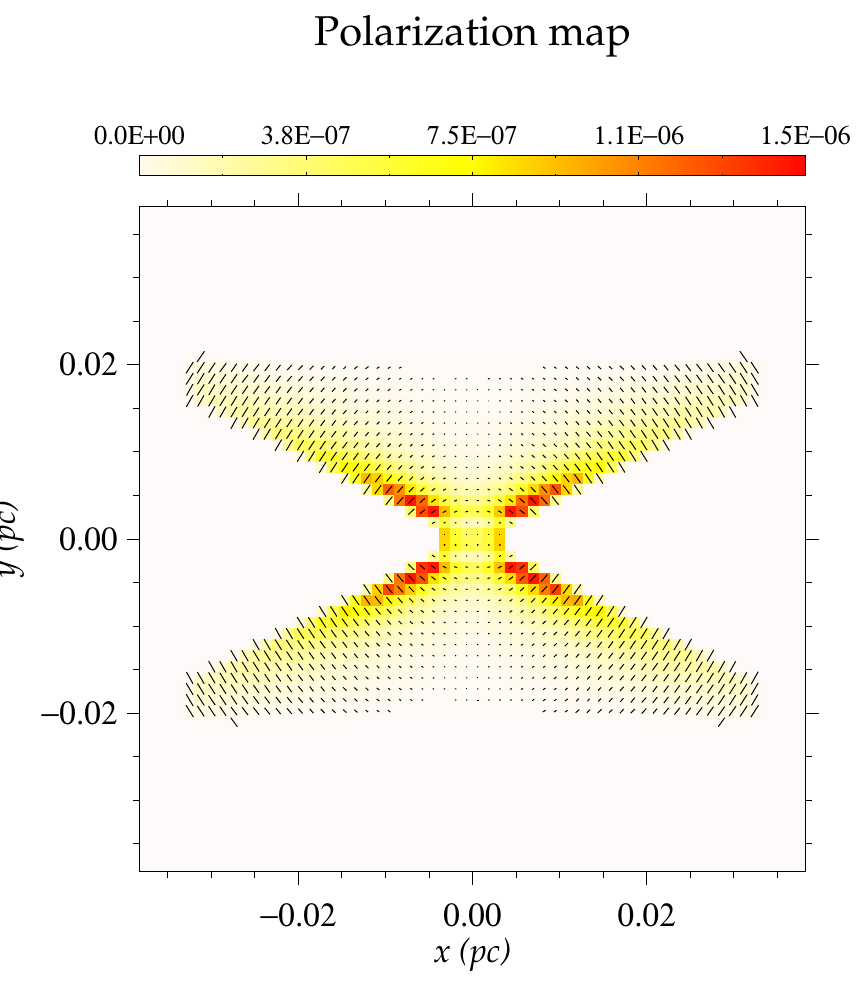}
   \caption{Modelled images of the $PF/F_{*}$ for an electron-filled,
            scattering outflow as presented in Fig.~\ref{Fig1.1}.
            The polarised flux, $PF/F_{*}$, is colour-coded and
            integrated over the wavelength band.
            \textit{Top}: image at $i \sim 9^\circ$;
            \textit{middle}: $i \sim 61^\circ$;
            \textit{bottom}: $i \sim 89^\circ$.}
  \label{Fig1.3}%
\end{figure}

The corresponding polarisation images are shown in Fig.~\ref{Fig1.3}. The 60~$\times$~60 spatial bins compile the polarisation properties integrated across 
the whole wave band of 2000 -- 8000~\AA. Each pixel is labelled by its projected coordinates $x$ and $y$ (in parsecs) with respect to the centre of the model 
space. The polarised flux $PF/F_{*}$ is colour-coded and the orientation and length of a vector drawn in black at the centre of each pixel represent the $\psi$ 
and $P$ values, respectively. The degree of polarisation is maximal for a full-length vector (size of a spatial bin); the length variation is proportional
to $P$.

Inspection of the polarisation images is helpful to understand the net polarisation properties as a function of the viewing angle: at a polar viewing direction 
(Fig.~\ref{Fig1.3}, top), the central source irradiates the outflow funnel, which causes a weak polarised flux in the vicinity of the model centre. However, 
the spatial distribution of polarisation position angles remains close to being symmetric with respect to the centre and therefore the resulting net polarisation 
at a low viewing angle is weak. Furthermore, the polarisation is strongly diminished by the diluting flux coming from the continuum source that is directly visible 
at low inclinations. The resulting polarisation position angle at $\psi = 90^\circ$ is determined by the geometry as the wind is rather flat ($\theta$ = 60$^\circ$) 
and therefore favours scattering close to the equatorial plane. Equatorial scattering produces parallel polarisation and was suggested early-on to play an important 
role in producing the polarisation properties of type-1 AGN \citep{Antonucci1984}.

At $i = 61\degr$ (Fig.~\ref{Fig1.3}, middle) the line-of-sight passes through the extended winds. The net polarised flux now comprises a strong component seen in 
transmission. While the wind base is optically thin, the scattering optical depth along the line-of-sight is significant and the latter thus contributes strongly 
to the net polarisation. The interface between the inner funnel of the torus and the extended winds is traced by a sharp gradient in polarised flux. The inner (outer) 
surfaces of the upper (lower) part of the outflow are visible in reflection, but due to their low optical depth in the poloidal direction the scattering is inefficient 
and does not produce much polarised flux.

Finally, the equatorial view of the model at $i = 89\degr$ (Fig.~\ref{Fig1.3},bottom) allows us to recover the edge-on morphology of the system. The polarised 
flux emerging from the wind base and its extensions is significantly higher than at the intermediate view. The boundaries of the extended winds trace out an X-shaped 
structure that extends far out and produces strong polarisation. These regions have a higher polarisation efficiency because they present a significant optical depth 
along the line of sight and favour a scattering angle around $90\degr$ with respect to the continuum source.

The polarisation position angle $\psi$ is equal to 90$^\circ$ for all viewing angles, indicating that the polarisation angle is oriented parallel to 
the symmetry axis of the system.\\

We remark that adopting the exact geometry suggested in \citet{Elvis2000} for a pure electron-scattering medium underestimates the observed optical 
polarisation percentage of type-1 and type-2 AGN. The wavelength-independent continuum polarisation induced by the WHIM produces a polarisation position 
angle $\psi = 90\degr$ at all viewing angles, whereas in type-2 AGN the resulting polarisation should be perpendicular. It is then necessary to look at 
more realisations of the model for slightly different parametrisations.

\subsubsection{Exploring different bending and opening angles of the wind}
\label{Sect2.2}

\begin{figure*}
    \begin{center}
      \begin{tabular}{cc}
	\includegraphics[trim = 10mm 0mm 0mm 0mm, clip, width=9cm]{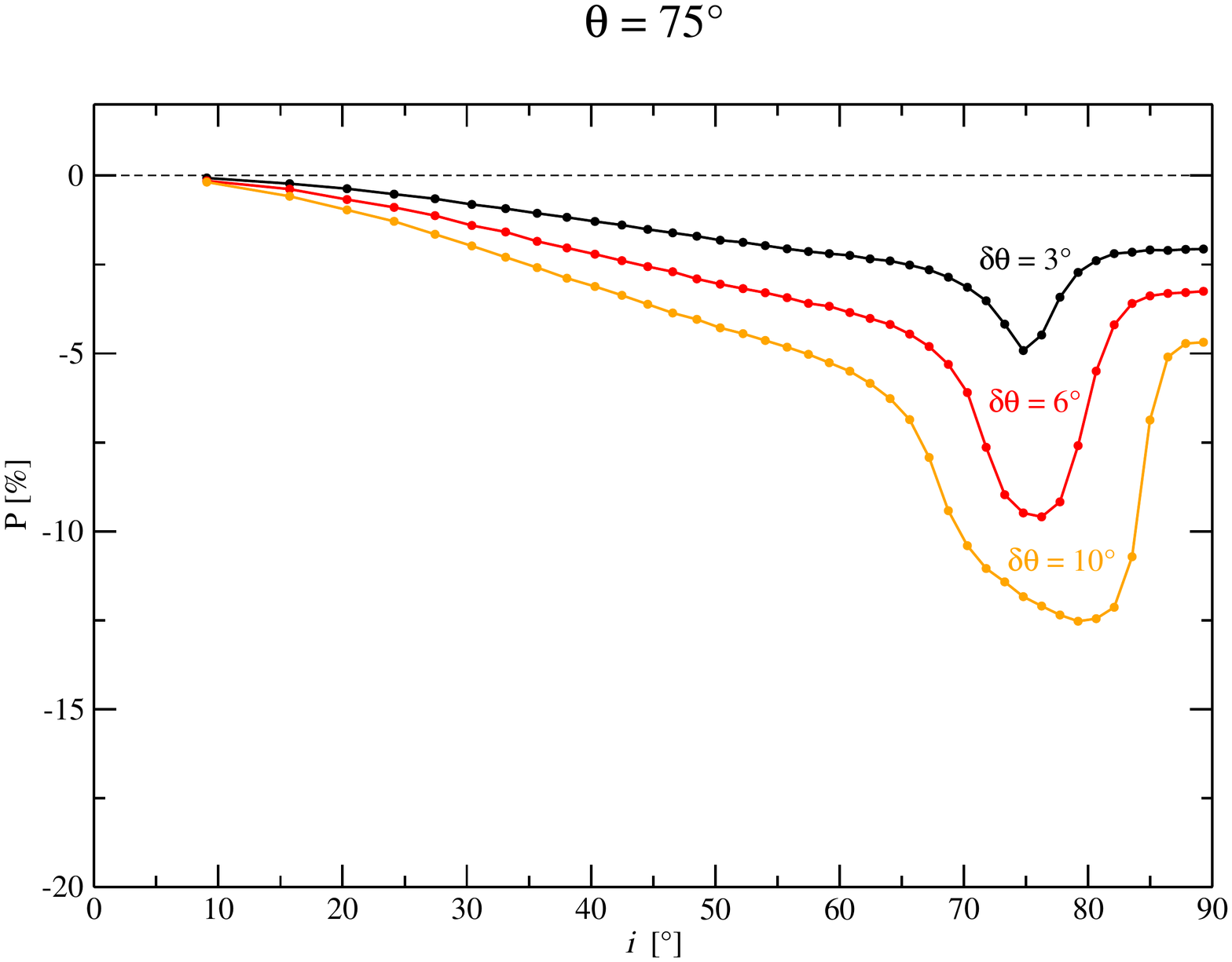} & \includegraphics[trim = 10mm 0mm 0mm 0mm, clip, width=9cm]{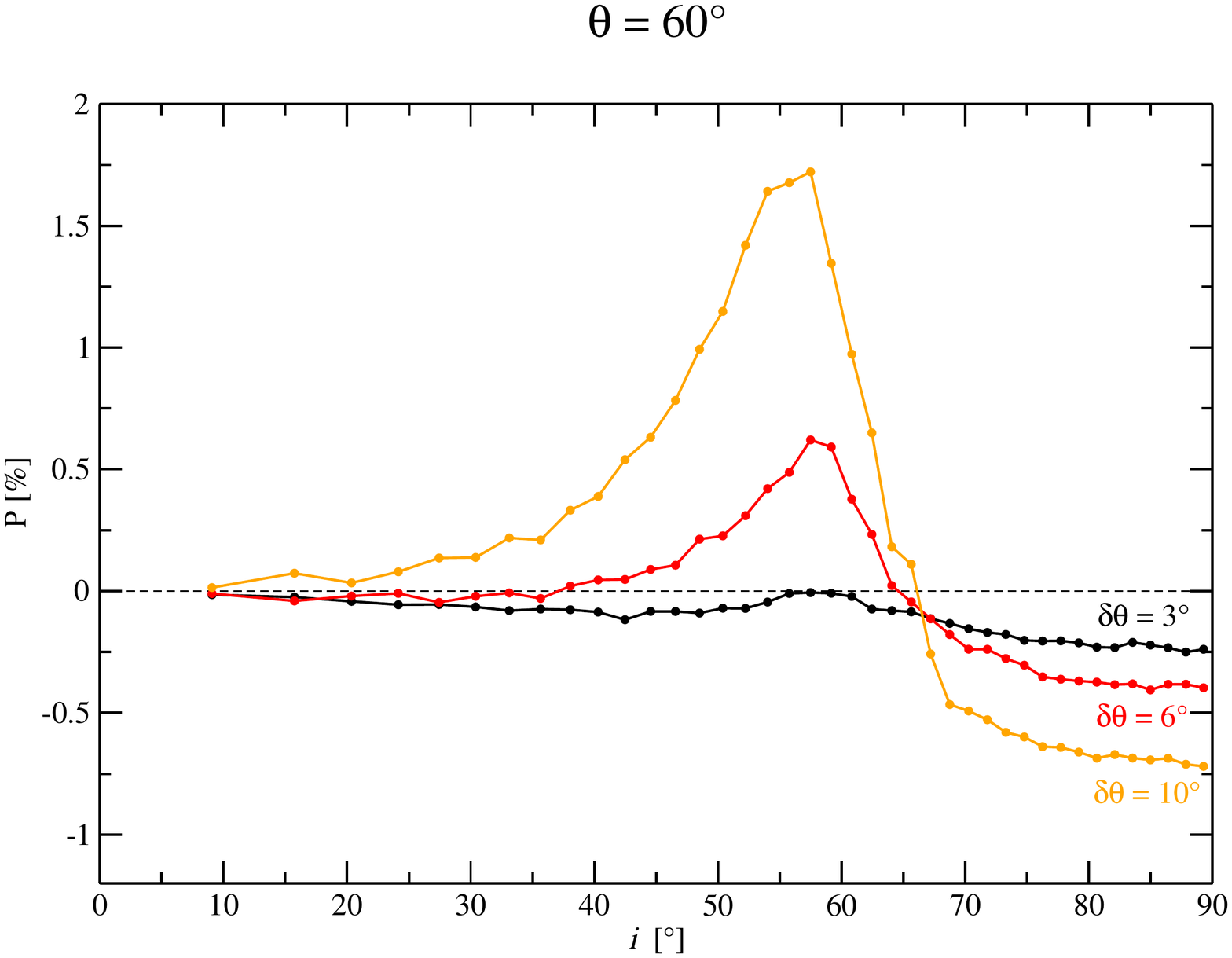} \\
	\includegraphics[trim = 10mm 0mm 0mm 0mm, clip, width=9cm]{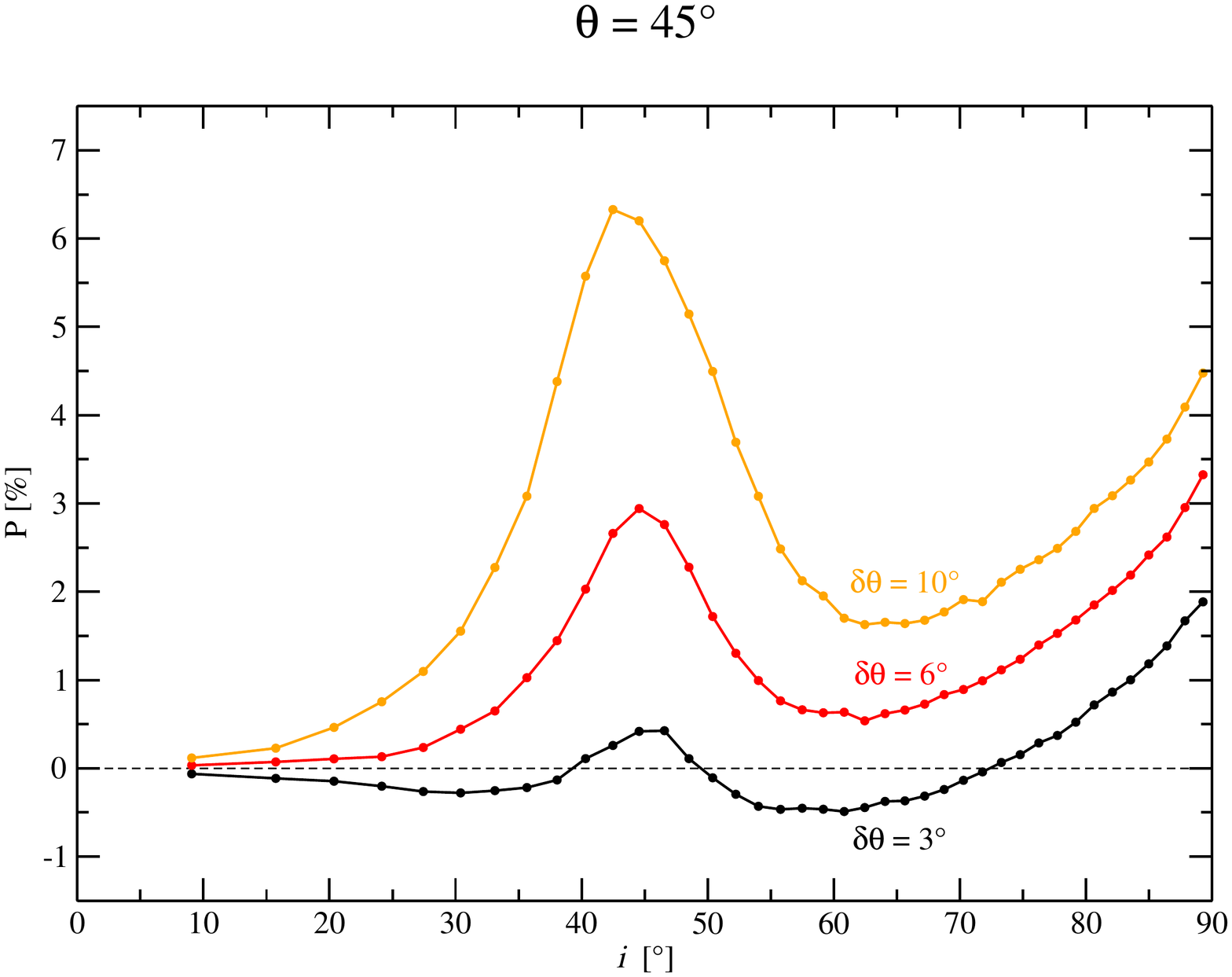} & \includegraphics[trim = 10mm 0mm 0mm 0mm, clip, width=9cm]{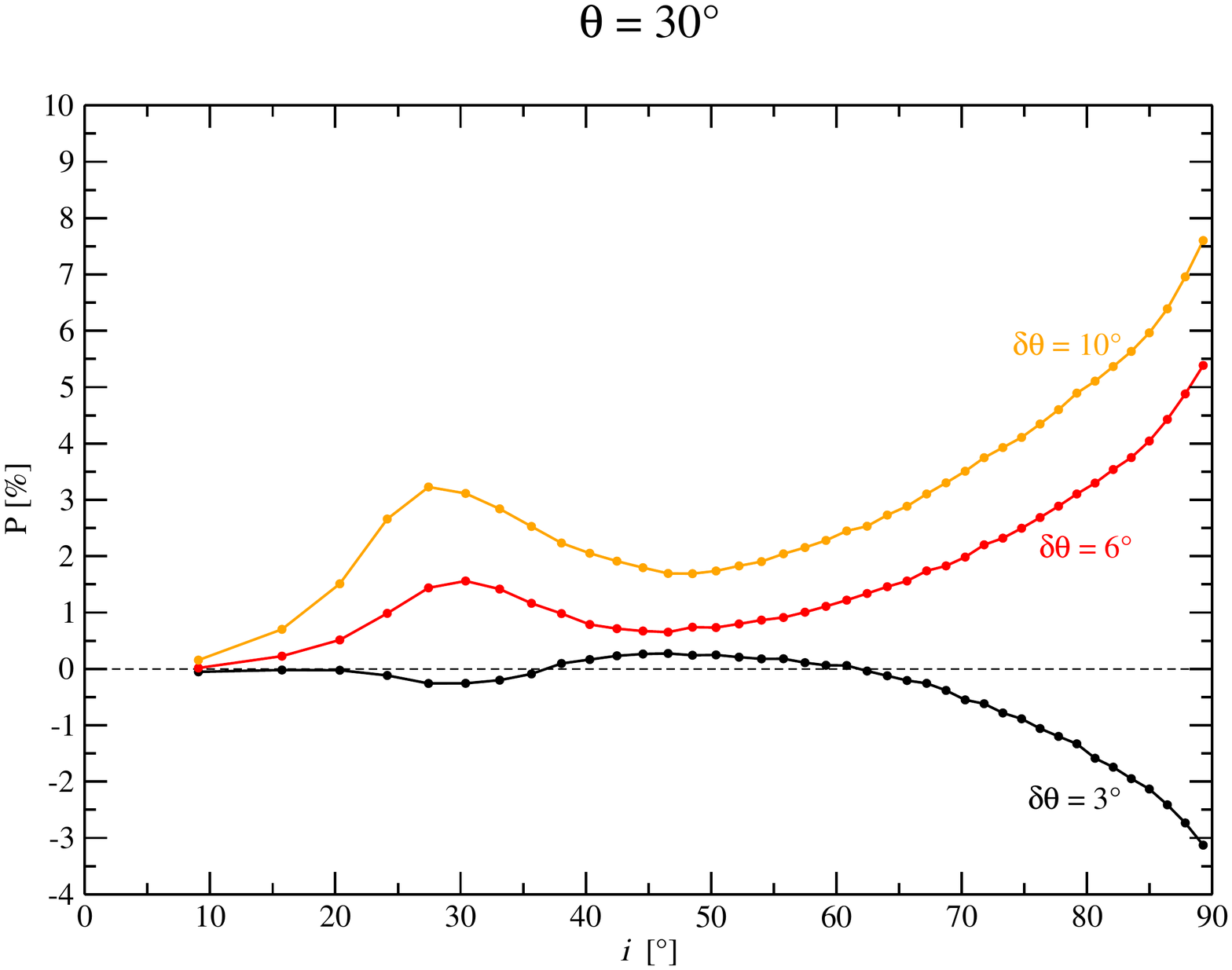} \\
      \end{tabular}
    \end{center}
    \caption{Investigating the polarisation response of the quasar's structure
	       according to the observer's viewing angle $i$,
	       at four different wind bending angles $\theta$ and three outflows 
	       opening angle $\delta\theta$ (black : $\delta\theta$ = 3$^\circ$ -- 
	       red : $\delta\theta$ = 6$^\circ$ -- orange : $\delta\theta$ = 10$^\circ$).
	       \textit{Top-left}: $\theta$ = 75$^\circ$;
	       \textit{top-right}: $\theta$ = 60$^\circ$;
	       \textit{bottom-left}: $\theta$ = 45$^\circ$;
	       \textit{bottom-right}: $\theta$ = 30$^\circ$.}
     \label{Fig2.1}%
   \end{figure*}

We continue our investigation by exploring different angles $\theta$ and $\delta\theta$ of the wind. To obtain a more narrow cone with respect to the default 
parametrisation used in Sect.~\ref{Sect2.1.1}, it is necessary to lower the bending angle $\theta$ while the wind extensions become thicker when increasing the half-opening 
angle $\delta\theta$. We systematically explore the response in polarisation for a range of both angles and summarise the results in Fig.~\ref{Fig2.1}. Since the 
polarisation of electron scattering is wavelength-independent we now plot the wavelength-integrated polarisation degree as a function of the viewing angle. Note that 
we apply a sign convention for $P$ that expresses the orientation of the polarisation position angle: for positive $P$, the $\vec E$-vector is perpendicular to the wind 
axis, while for negative $P$ the polarisation is parallel (intermediate orientations of the net polarisation are impossible due to symmetry reasons).

For $\theta = 75\degr$ the model produces polarisation degrees that are low at small viewing angles and moderate, up to 15~\%, for higher inclinations. The normalisation of 
$P$ rises with $\delta\theta$. There is no dichotomy for the polarisation angle as the $\vec E$-vector is always parallel to the projected axis. The case of $\theta = 60\degr$ 
with $\delta\theta = 3\degr$ is discussed in the previous section. It turns out that increasing $\delta\theta$ can change the dichotomy of the polarisation position angle and
even produce three or four different regimes at low, intermediate and large viewing angles. The polarisation then turns out to be parallel for a face-on and edge-on view and 
perpendicular along a line-of-sight through the wind. Further narrowing the bending angle to $\theta = 45\degr$ increases the trend to have multiple regimes for the polarisation 
angle that now occur even for the smallest opening angle of $\delta\theta = 3\degr$. It also reinforces the polarisation degree for parallel polarisation until an optimum is 
reached and the trend becomes inverted (see the case of $\theta$ = 30$\degr$).

An important motivation for the model by \citet{Elvis2000} was given by the observed number distribution of quasars with narrow, ionised absorption lines in the UV and 
X-ray band versus BAL- and non-NAL objects. The BAL objects are assumed to be seen along the conical wind, the NAL objects on a line of sight that crosses the optically 
thin wind base. Quasars with NALs are type-1 AGN and therefore preferentially have a parallel polarisation in the optical. Interestingly, there is a parameter range for 
which our modelling allows for the existence of parallel polarisation at low inclinations (non-NAL, type-1 AGN with low parallel polarisation) and high-inclinations (NAL, 
type-1 AGN with significant parallel polarisation). In between these inclinations the line of sight points toward a BAL objects with considerable perpendicular 
polarisation. Immediate observational tests for the number distribution of BAL, NAL, and non-NAL quasars can be performed to test the possible angles $\theta$ and 
$\delta\theta$.\\

An interesting challenge to the wind model by \citet{Elvis2000} is the presence of obscured, type-2 AGN without broad absorption lines and a ubiquitous perpendicular polarisation. 
These objects may be accounted for by adding dust shielding at certain inclinations to the model. The dust may exist on the far-side of the wind where the temperature drops below 
the sublimation temperature. We are going to investigate to such scenarios in the following.

\subsection{Dust in the wind}
\label{Sect2.3}

The structure for quasars analysed here may also account for the extended narrow line regions found at greater distances from the central engine. They are supposed to be the 
extensions of the ionised outflow, beyond the sublimation radius where dust can survive. However, the presence of dust originating close to the accretion disc remains theoretically 
possible according to \citet{Czerny2012}, who demonstrated that dust can be formed in accretion disc atmospheres. Dust rises similarly to the electronic disc-born wind, but is soon evaporated 
by the central irradiation source \citep{Elvis2012}. Knowing that a large fraction of the thermal, infrared emission observed in low-luminosity AGN (LLAGN) is associated with the presence 
of dusty matter \citep{Edelson1987}, \citet{Elvis2000} suggested that a part of the BAL flow may be obscured by a dust region that is still close to the active nucleus. It is a 
natural explanation of the absence of BAL in Seyfert-like galaxies and fits in with the observed presence of dusty NLR clouds at larger distances from the irradiation source.

\begin{figure*}
   \centering
      \includegraphics[trim = 0mm 145mm 0mm 40mm, clip, width=17cm]{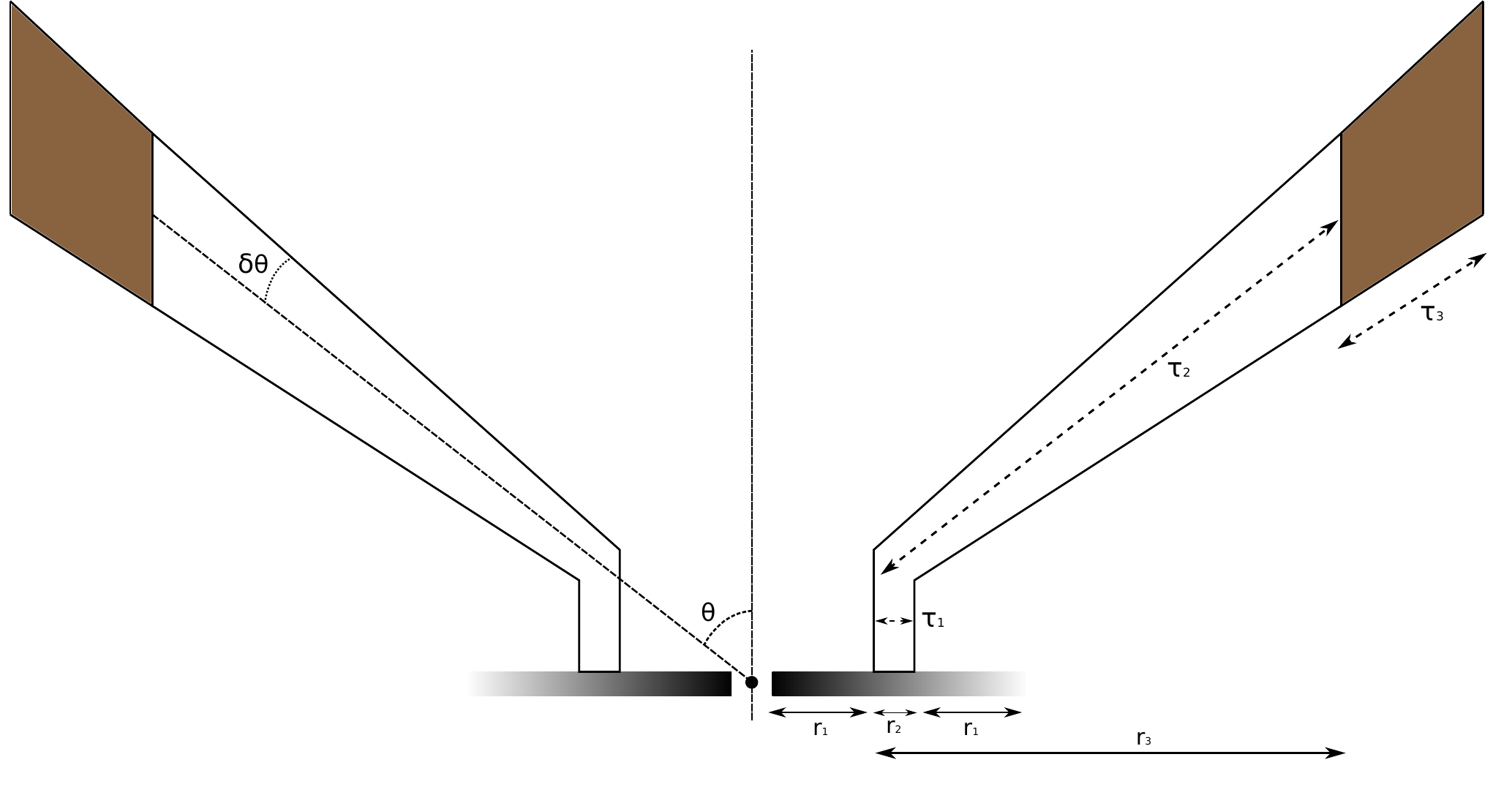}
      \caption{Investigating the presence of dust in low luminosity objects. 
	       The brownish medium represents the dusty medium,
	       preventing the BAL from being seen in Seyfert-2 galaxies.}
     \label{Fig4.1}
   \end{figure*}

We simulate the BAL obscuring wind using our previous WHIM model (see Sect.~\ref{Sect2.1}) and adding two cylindrically shaped extinction regions to the outer edge of the flow. The dusty 
wind originates at r$_3$ = 0.032~pc (0.8~$T_{\tau, \rm K}$) and is of moderate optical depth ($\tau_3 \sim$ 1) in order to allow the radiation to partially escape along intermediate viewing 
angles. The model is summarised in Fig.~\ref{Fig4.1}). The dust composition and grain size distribution represents an average dust type of our Milky Way \citep[see][]{Mathis1977,Wolf1999,Goosmann2007}.

\begin{figure}
   \centering
      \includegraphics[trim = 10mm 0mm 0mm 0mm, clip, width=10cm]{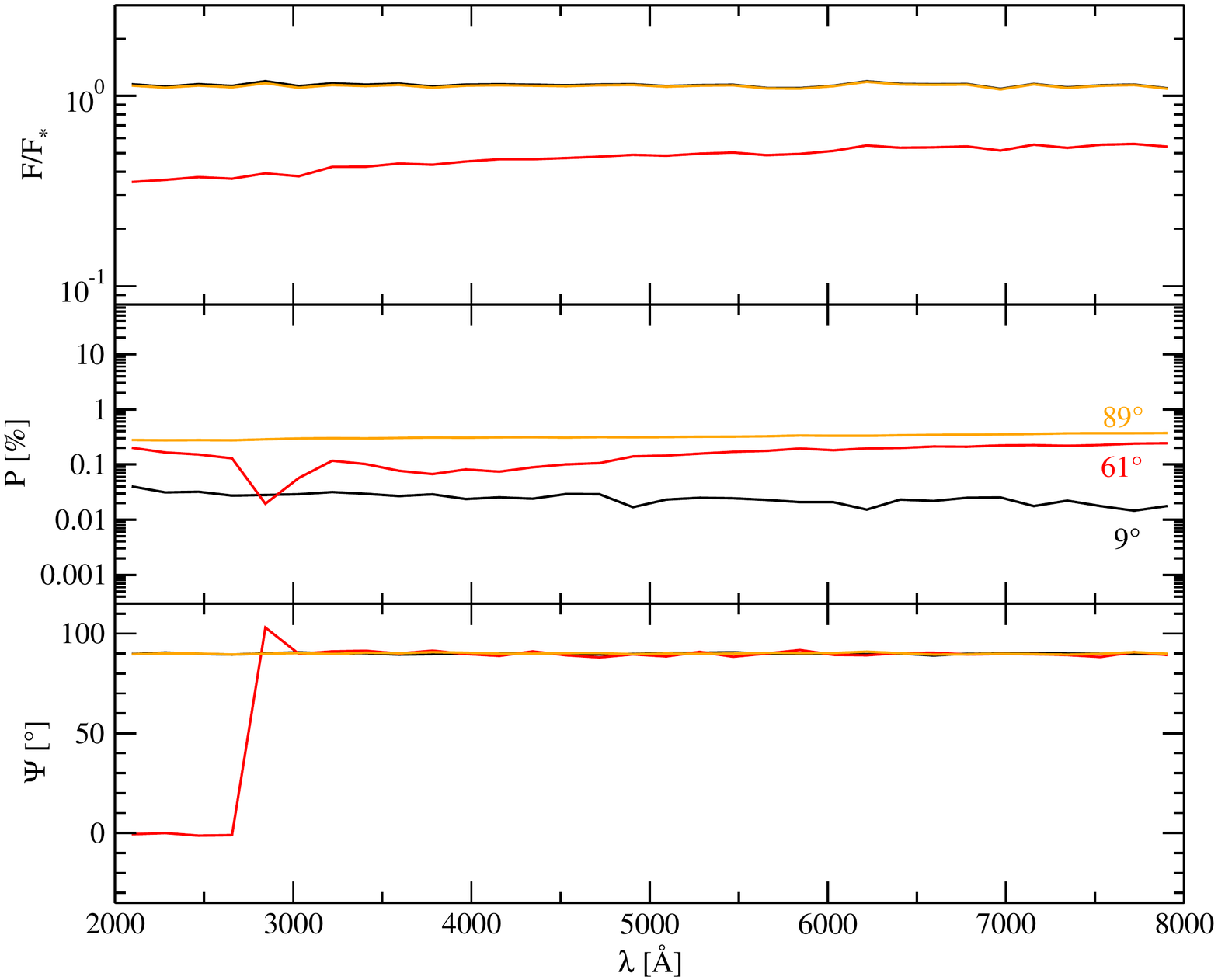}
      \caption{Modelling an obscured structure for quasars outflow
	       where BAL are seen through a dusty medium of $\tau_{dust} \sim 1$.
	       The model, presented in Fig.~\ref{Fig4.1}, is
	       seen at different viewing inclinations, \textit{i} 
	       (black : 9$^\circ$ -- red : 61$^\circ$ -- orange : 89$^\circ$).
	       \textit{Top}: the fraction, $F$/$F_{\rm *}$ of the central flux;
	       \textit{middle}: polarisation, \textit{P};
	       \textit{bottom}: polarisation angle, $\psi$.}
     \label{Fig4.2}
   \end{figure}

The resulting total flux spectra for polar and equatorial viewing angles (Fig.~\ref{Fig4.2}) are similar to those for the WHIM model (Fig.~\ref{Fig1.2}), indicating that Thomson scattering 
is the predominant reprocessing mechanism occurring at these inclinations. The impact of the dust on $F$/$F_{\rm *}$ remains marginal except when the observer's line-of-sight crosses the 
outflowing direction ($i = 61^\circ$). The flux then becomes wavelength-dependent and increases towards the red, but with respect to the pure WHIM case it does not suffer from heavy 
absorption due to the moderate opacity of the dusty wind. The impact of the dust clouds is particularly visible in the percentage of polarisation. At polar inclinations, $P$ remains 
below 0.05~\% and is nearly wavelength-independent. At equatorial inclinations, $P$ is at its maximum ($\sim$ 0.3~\%) and again wavelength-independent. However, at an intermediate 
inclination, the escaping radiation crosses the dust atmosphere and presents the characteristic wavelength-dependent polarisation signal. The net polarisation degree is higher than 
for the pure WHIM model as photons have to scatter into the line-of-sight; $P$ drops from 0.2~\% to 0.02~\% in the blue part of the spectrum and then rises back to 0.2~\% in the red. 
This peculiar dip is connected to a switch of the polarisation position angle, from 0$^\circ$ to 90$^\circ$, as the Mie scattering phase function becomes more forward-dominated and the 
overall scattering geometry thus changes towards shorter wavelengths. In this regard, the wavelength-dependent impact of dust scattering onto the polarisation position angle may serve 
as a tracer of the optical thickness of the obscuring region along the outflowing direction. At all other inclinations and wavelengths of the system the polarisation 
position angle is parallel to the system axis. 

Thus, adding dust to the distant parts of the wind helps $P$ to remain globally low, below 0.3~\% (which is in agreement with the observations of non-BAL radio-quiet 
AGN \citep{Berriman1990} where a mean $P$ value of 0.5~\% was found) but does not help to produce the dichotomy of the polarisation position angle in the optical.

\subsection{A two-phase medium}
\label{Sect2.5}

An important test of the structure for quasars is to reproduce the polarisation position angle dichotomy observed in AGN \citep{Antonucci1983,Antonucci1984}; our results in 
Sect.~\ref{Sect2.2} have proven that a pure WHIM is yet inadequate to explain the switch in polarisation position angle between type-1 and type-2 AGN. In the context of Seyfert-2 AGN, 
the additional presence of dust close to the equatorial plane is suggested as a physical possibility \citep{Czerny2012}. Shielded from the full continuum by the WHIM, dust can rise 
from the disc and survive long enough to impact the radiation across a significant solid angle around the equatorial plane. Such an absorbing medium would be less ionised than the 
high ionisation BEL region, giving rise to low-ionisation line emission, such as in NGC~5548 where the Mg{\sc II} lines are thought to arise from several light-months away from the 
irradiating source \citep{Elvis2000}. However, the equatorial dust should also suppress the parallel polarisation signal at intermediate and equatorial viewing angles.

In the following, we investigate the polarisation properties of a two-phase structure, where the highly ionised inner flow self-shields radially a cold absorbing flow arising farther away 
from the continuum source. Such a medium should be optically thin at some viewing angle in order to allow for the detection of NAL signatures above the equatorial dust lane, but 
not too optically thin to prevent the production of parallel polarisation at other inclinations.

   \begin{figure*}
   \centering
      \includegraphics[trim = 0mm 145mm 5mm 40mm, clip, width=18cm]{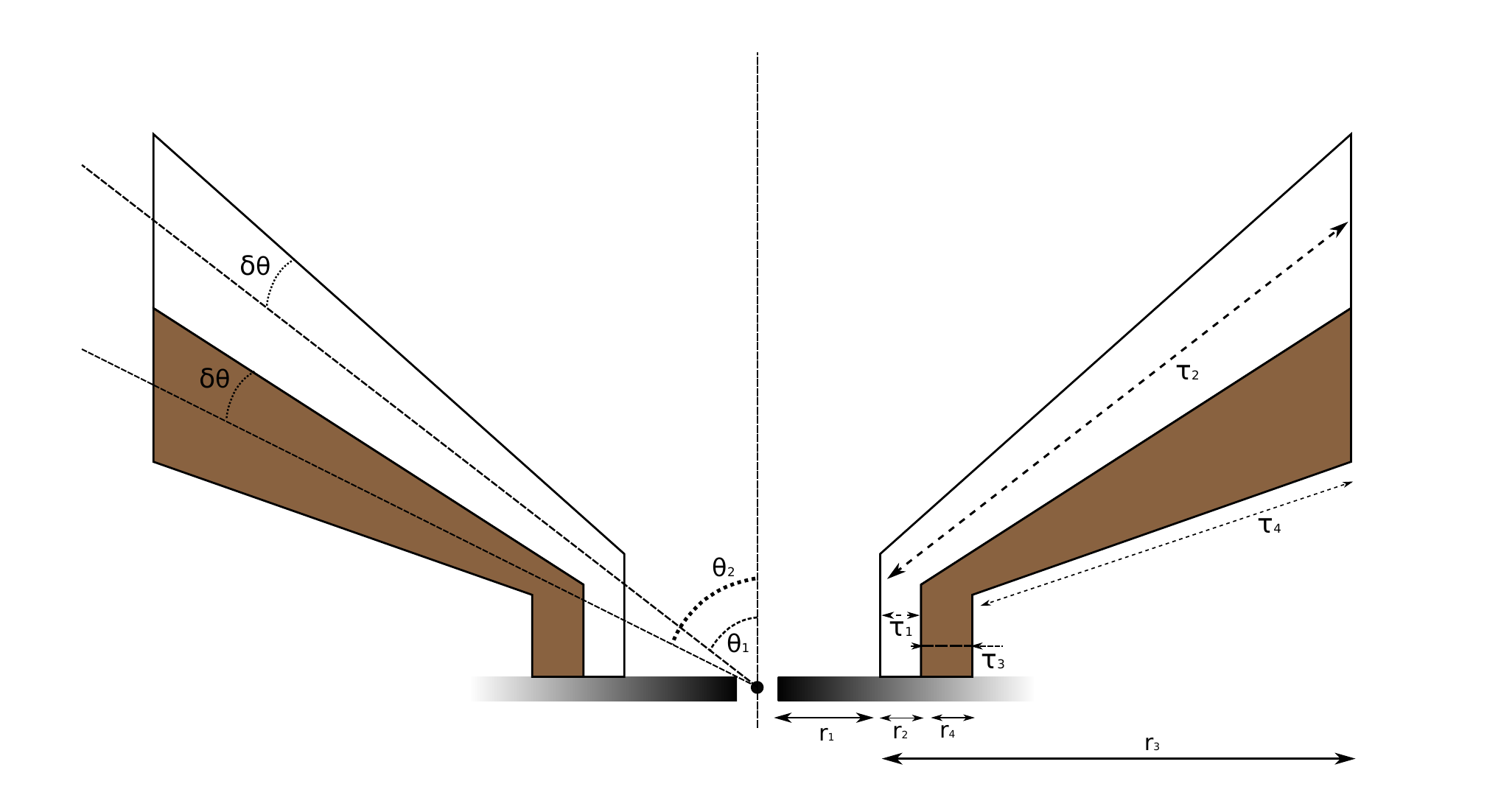}
      \caption{A two-phased medium where the cool WHIM outflow is composed of dust.}
     \label{Fig5.1}
   \end{figure*}

\subsubsection{Modelling two-phase outflows}
\label{Sect2.5.1}

We adopt the morphology proposed by \citet{Elvis2000} for NGC~5548 to explain the high/low ionisation BEL regions. Of course, dust would not survive in the BLR, so the dusty wind we 
define here actually needs to be farther away from the centre or remain cold by efficient shielding in the WHIM and BLR. The model geometry is presented in Fig.~\ref{Fig5.1} and 
we fix $\tau_1 \sim$~0.02, $\tau_2 \sim$~2, $\tau_3 \sim$~40 and $\tau_4 \sim$~4000 ($\tau$ being integrated over the 2000 - 8000~\AA~band). The bending angle of the WHIM, 
$\theta = \theta_1$, is equal to 60$^\circ$ and $\theta_2$ = $\theta_1$ + 2$\delta\theta$. Similarly to our previous modelling, $\delta\theta$ = 3$^\circ$. According to 
\citet{Elvis2000}, r$_1$ = 0.0032~pc, r$_2$ = 0.00032~pc, r$_3$ = 0.032~pc and r$_4$ = 0.00608~pc (1.88$\times$10$^{16}$~cm = 12700~$R_{\rm G}$ = 0.15~$T_{\tau, \rm K}$ = 
1.74~$R_{\rm BLR}$). 

To consider a more physical emission geometry, we replace the radiating disc by a point-like source. With the flux intensity decreasing like $r^{-3}$ from the centre, most of 
the radiation comes from the inner part of the emitting disc. When comparing different realisations of the modelling, we find, however, that the size of the emission region does 
not have a major impact on the spectropolarimetric results. We now also include the fact that the inner accretion flow allows for Thomson scattering and we define a geometrically 
flat, flared-disc region, with optical depth $\tau \sim$~1 along the equator. This reprocessing region was proposed early on by \citet{Antonucci1984} to explain the presence of 
parallel polarisation in type-1 AGN, and a complete description of its polarisation properties can be found in \citet{Goosmann2007}. The flared morphology is in agreement with 
the theory of accretion discs \citep{Shakura1973} and will strengthen the production of parallel polarisation, with a maximum polarisation degree of up to a few percent \citep{Marin2012a}.
However, we checked that it only marginally contributes to the production of parallel polarisation, the bulk of it emerging from the double wind structure.

   \begin{figure}
   \centering
      \includegraphics[trim = 10mm 0mm 0mm 0mm, clip, width=10cm]{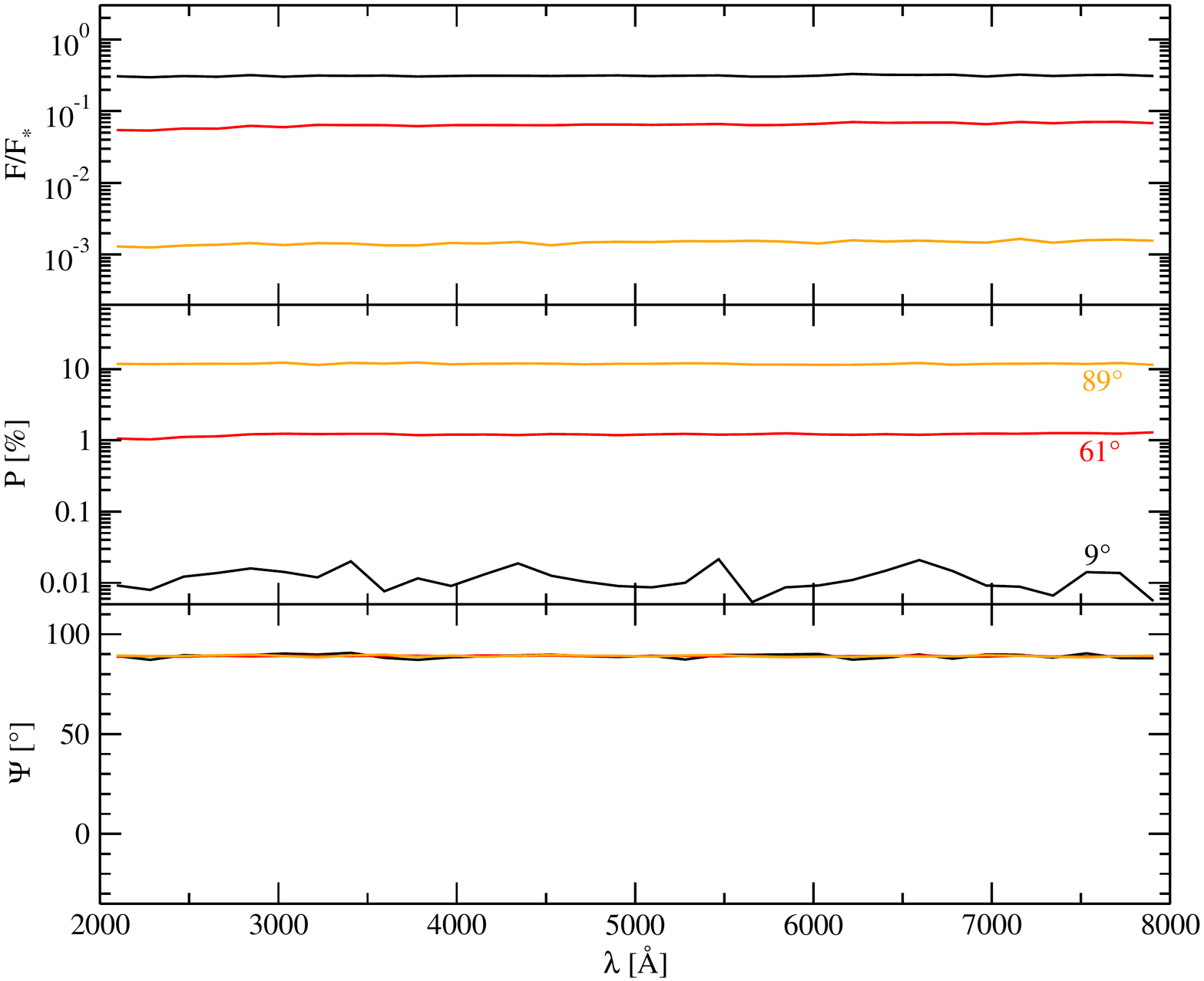}
      \caption{Modelling a two-phased quasars outflow where the cool WHIM outflow
	       is composed of an optically thick dust medium,
	       using an uniform model such as in Fig.~\ref{Fig5.1},
	       seen at different viewing inclinations, \textit{i} 
	       (black : 9$^\circ$ -- red : 61$^\circ$ -- orange : 89$^\circ$).
	       \textit{Top}: the fraction, $F$/$F_{\rm *}$ of the central flux;
	       \textit{middle}: polarisation, \textit{P};
	       \textit{bottom}: polarisation angle, $\psi$.}
     \label{Fig5.2}
   \end{figure}

The spectropolarimetric modelling results are plotted in Fig.~\ref{Fig5.2}. Interestingly, the polar viewing angle shows a maximal and wavelength-independent flux in comparison with 
other inclinations, as expected from observations of radio-quiet, type-1 AGN. The impact of dust is visible by absorption and anisotropic scattering processes in intermediate and 
equatorial inclinations, where the obscured flux is a hundred times lower than at polar lines-of-sight. The percentage of polarisation is minimal for polar inclinations ($P \approx $ 0.01~\%) 
and maximal for equatorial viewing angles ($P \approx $ 10~\%). Along the direction of the outflow, the polarisation ranges around $1~\%$, is wavelength-independent, and oriented parallel to 
the projected axis of the model. The position angle is the same at other inclinations and the expected polarisation dichotomy is still not reproduced. Therefore, the two phase model needs 
to be further refined.\\

So, as expected from observations, brightest fluxes are produced for unobscured, type-1 viewing angles but the model fails to reproduce neither acceptable levels of polarisation 
nor the observed polarisation dichotomy. Some morphological modifications must be considered to aim for polarisation degrees of the order of 1~\% and 5~\% at type-1 \citep{Smith2002} 
and type-2 \citep{Kay1994} inclinations, respectively.

\subsubsection{Investigating different opening angles}
\label{Sect2.5.2}

Similarly to Sect.~\ref{Sect2.3}, we now explore a broader range of morphological parameters ($\theta$, $\delta\theta$) in order to discriminates between more realistic and less 
favourable realisations of the two-phase wind model. As seen from Fig.~\ref{Fig5.2}, the polarisation continuum is close to being wavelength-independent, thus we will also plot 
the wavelength-integrated polarisation degree as a function of the viewing angle. The sign convention is preserved from Sect.~\ref{Sect2.3}.

   \begin{figure*}
    \begin{center}
      \begin{tabular}{cc}
	\includegraphics[trim = 10mm 0mm 0mm 0mm, clip, width=9cm]{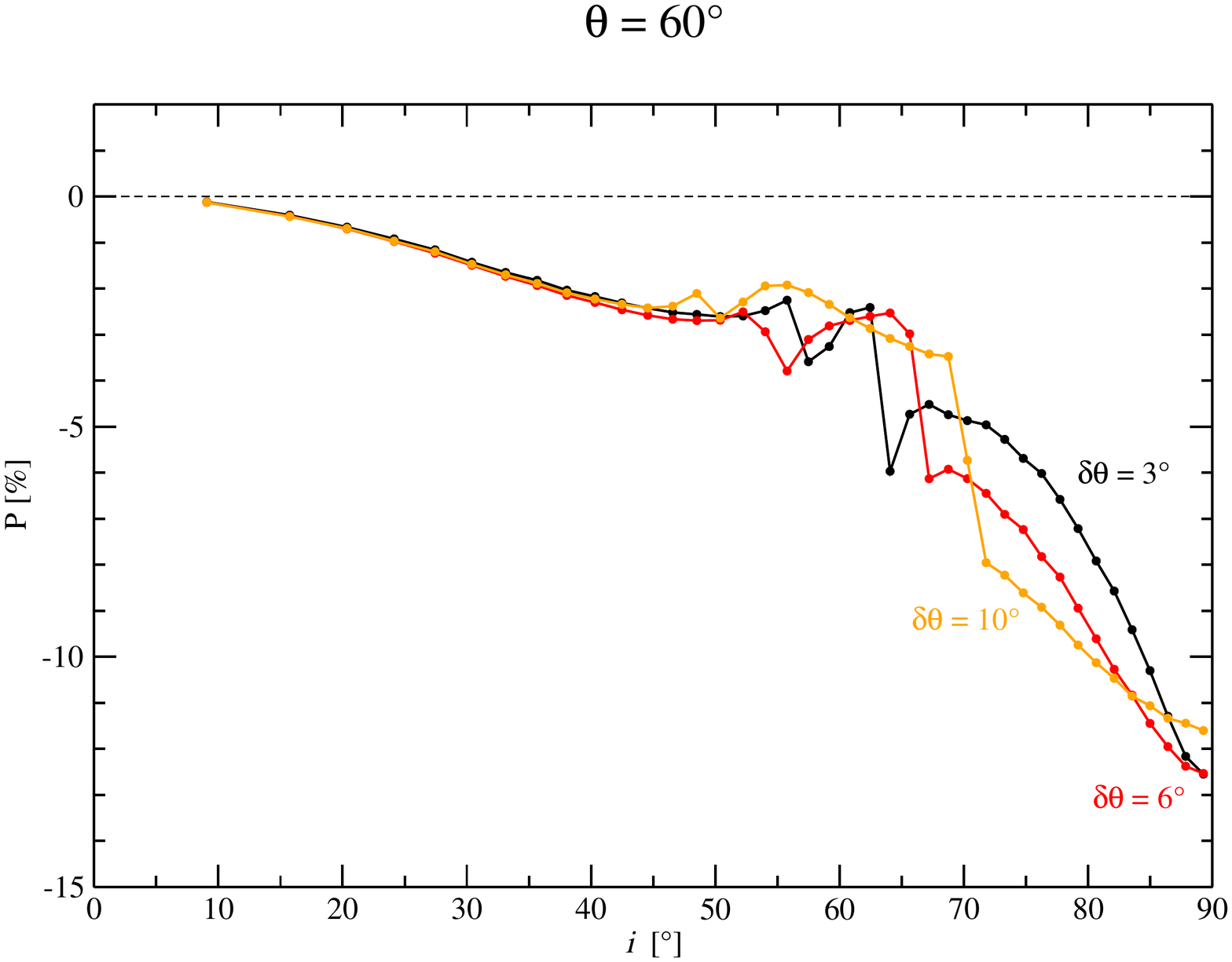} 
      \end{tabular}
      \begin{tabular}{cc}
	\includegraphics[trim = 10mm 0mm 0mm 0mm, clip, width=9cm]{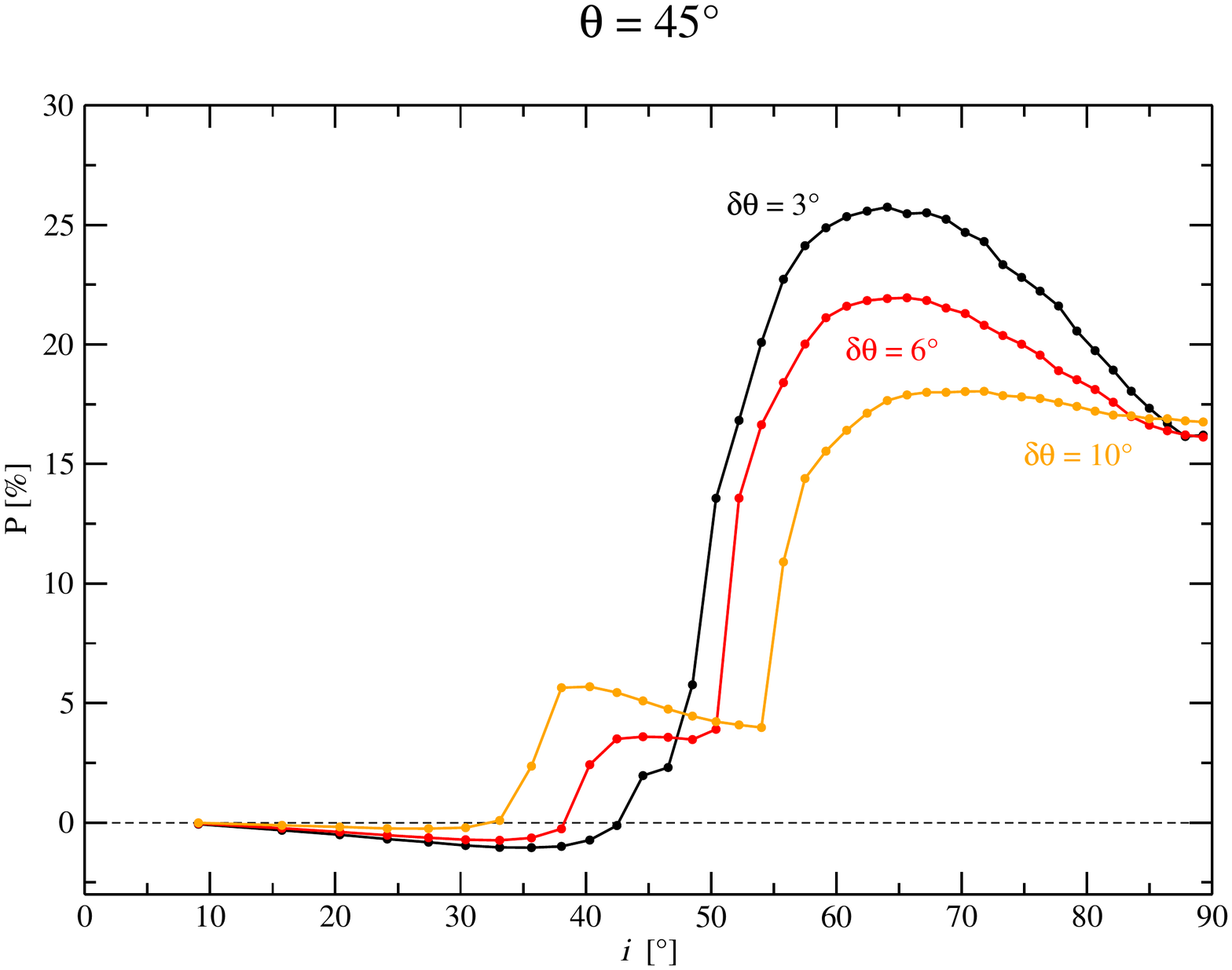} & \includegraphics[trim = 10mm 0mm 0mm 0mm, clip, width=9cm]{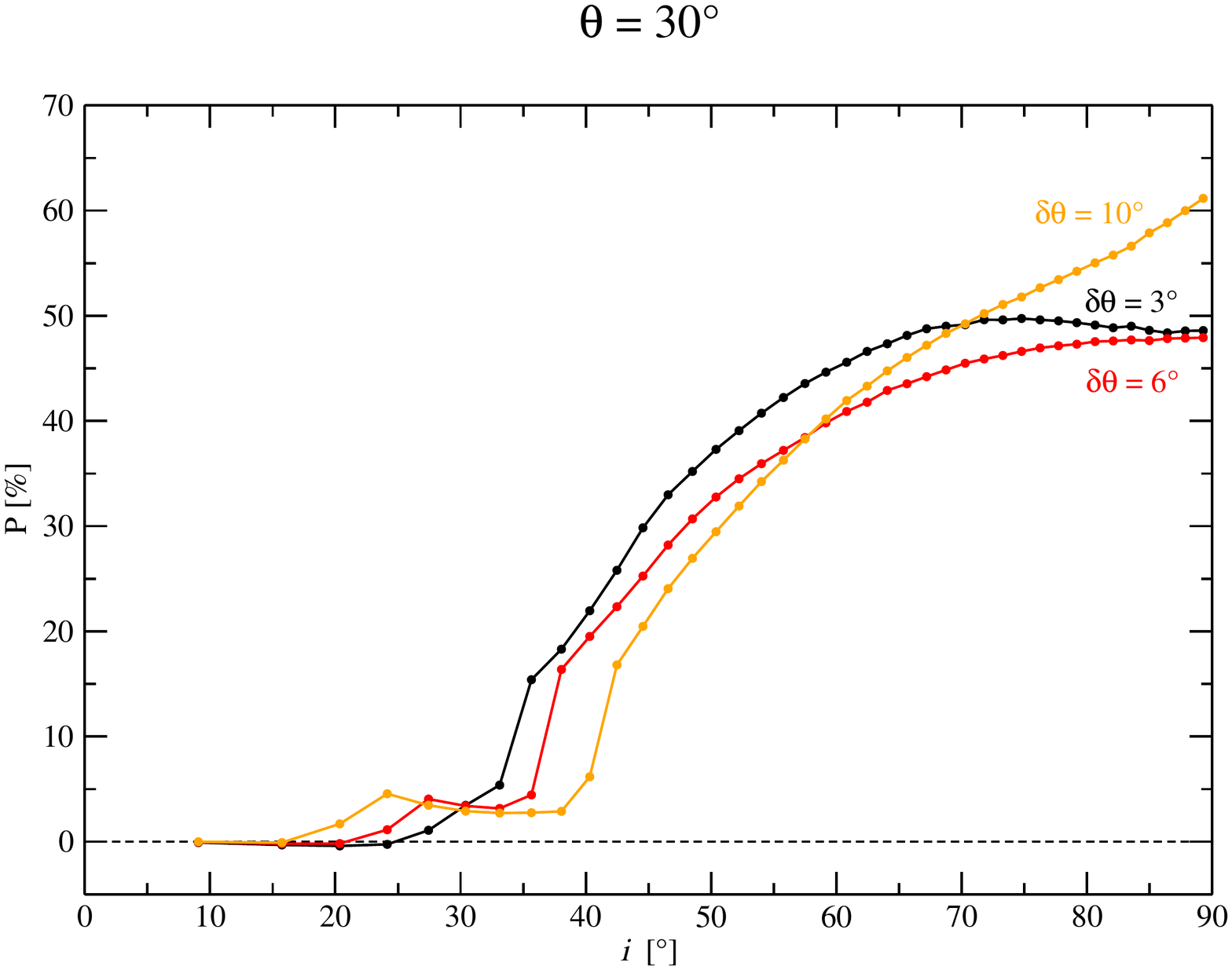} \\
      \end{tabular}
    \end{center}
    \caption{Investigating the polarisation response of the dual quasar's structure
	       according to the observer's viewing angle $i$,
	       at three different wind bending angles $\theta$ and three outflow's 
	       opening angle $\delta\theta$ (black : $\delta\theta$ = 3$^\circ$ -- 
	       red : $\delta\theta$ = 6$^\circ$ -- orange : $\delta\theta$ = 10$^\circ$).
	       \textit{Top}: $\theta$ = 60$^\circ$;
	       \textit{bottom-left}: $\theta$ = 45$^\circ$;
	       \textit{bottom-right}: $\theta$ = 30$^\circ$.}
     \label{Fig5.3}%
   \end{figure*}

For $\theta$ = 60$^\circ$ (Fig.~\ref{Fig5.3}), the model uniformly creates negative (parallel) polarisation for the whole range of half-opening angles, with moderate polarisation 
degrees at low inclinations, increasing up to 12~\% at equatorial viewing angles. The parametrisation is non compatible with the expected polarisation dichotomy. However, the case 
of $\theta$ = 45$^\circ$ is more interesting as it correctly reproduces the variation of the polarisation position angle between type-1 and type-2 inclinations. The transition 
between negative to positive polarisation is a function of the half-opening angle of the model, but is comprised between $i \sim$ 34$^\circ$ and $i \sim$ 42$^\circ$. 
Negative [positive] polarisation degrees reach up to 1.2~\% [25.6~\%] for the flattest flow geometry ($\delta\theta$ = 3$^\circ$) and the polarised flux tends to decrease when the 
observer's line-of-sight is reaching the equatorial plane ($P$ = 16.3~\%). Finally, for $\theta$ = 30$^\circ$, the polarisation dichotomy also appears but the regime transition 
occurs between $i \sim$ 15$^\circ$ and $i \sim$ 24$^\circ$, which is unrealistic according to the observed distribution of AGN types and individual observations of Seyfert-1 inclinations, 
such as for ESO~323-G077 ($i$ = 45.0$^\circ$, \citealt{Schmid2003}) or NGC~5548 ($i$ = 47.3$^{+7.6}_{-6.9}$, \citealt{Wu2001}). Moreover, the polarisation degree for models with 
$\theta$ = 30$^\circ$ can be as high as 62~\% at equatorial inclinations, a behaviour far from polarimetric observations of obscured AGN \citep{Kay1994}. \\

Our immediate conclusions are coherent with the ones derived from electron-filled outflowing models presented in Sect.~\ref{Sect2.3} : a model with a wind bending angle 
$\theta$ = 60$^\circ$ unsuccessfully reproduces the polarisation dichotomy, but a change in the flow geometry can solve this issue. As stated in Sect.~\ref{Sect2.3}, immediate 
observational tests for the number of BAL, NAL, and non-NAL quasars can be performed to test the possible angles $\theta$ and $\delta\theta$. In particular, for 
$\theta$ = 45$^\circ$, both the transition between parallel and perpendicular polarisation at acceptable inclinations and polarisation degree close to observed 
quantities are emerging (non-NAL quasars polarisation $\sim$ 1~\%, NAL-objects $\sim$ 16~\% and BALQSO's $\sim$ 25~\%). The transition regime between type-1 and type-2 
signatures appears to be below $i = 45^\circ$, which is in agreement with the inclination of borderline type-1 objects \citep{Schmid2003}.

\subsubsection{Exploring a range of dust optical depths}
\label{Sect2.5.3}

It remains a challenge to the double-wind model to reproduce the observed levels of polarisation. With respect to the results of Sect.~\ref{Sect2.5.2}, the polarisation degree 
should be reduced by a factor of five at type-2 inclinations, and by a factor of two at intermediate viewing angles. The cold, absorbing medium shielded by the WHIM may play a 
decisive role in this. 

\begin{figure*}
    \begin{center}
      \begin{tabular}{cc}
	\includegraphics[trim = 10mm 0mm 0mm 0mm, clip, width=9cm]{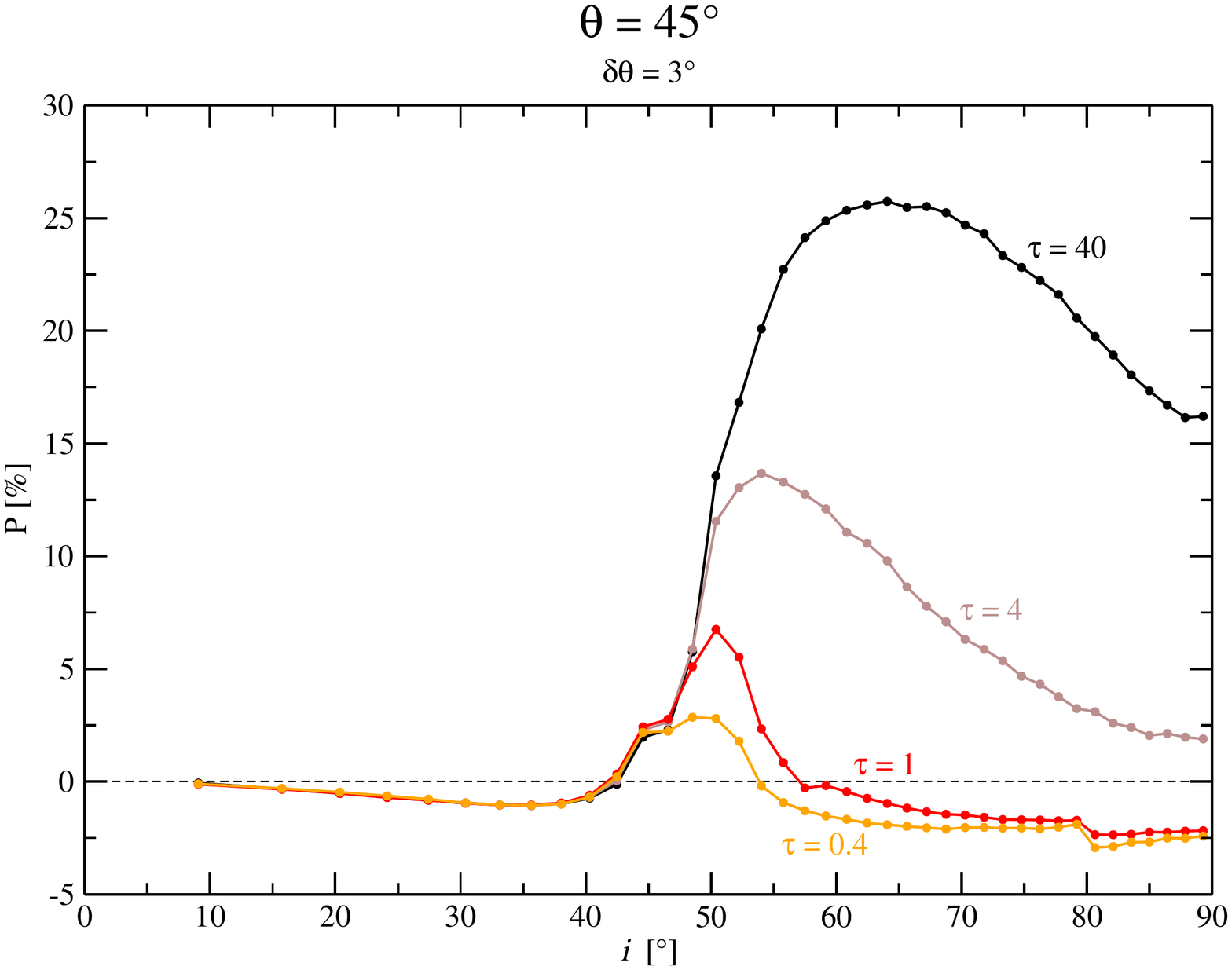} 
      \end{tabular}
      \begin{tabular}{cc}
	\includegraphics[trim = 10mm 0mm 0mm 0mm, clip, width=9cm]{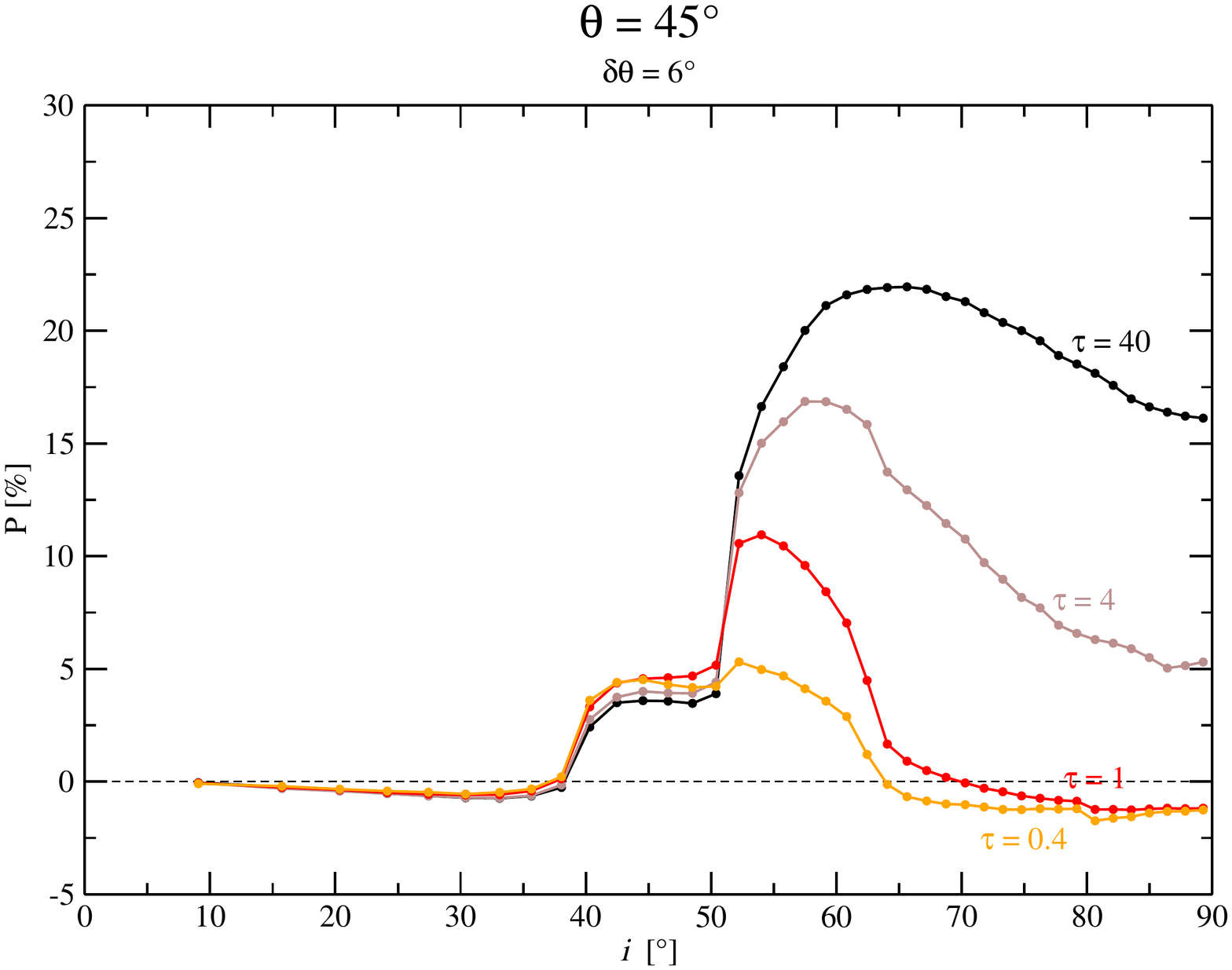} & \includegraphics[trim = 10mm 0mm 0mm 0mm, clip, width=9cm]{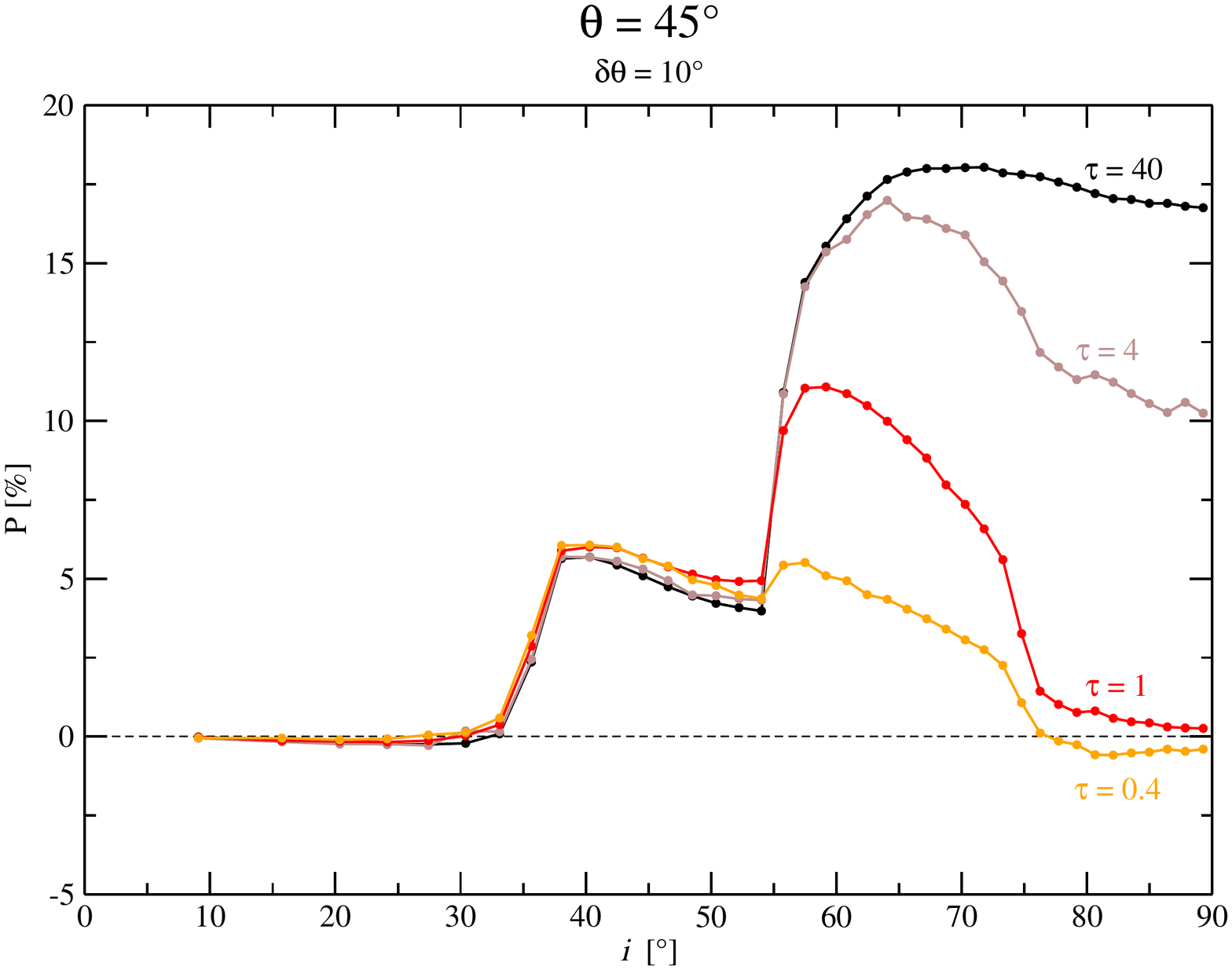} \\
      \end{tabular}
    \end{center}
    \caption{Investigating the polarisation response of the dual quasar's structure
	       with a bending angle $\theta$ = 45$^\circ$ according to the observer's viewing 
	       angle $i$, varying the outflow's opening angle $\delta\theta$ and the dust 
	       equatorial optical depth $\tau$ (black : $\tau_{dust}$ = 40 -- 
	       brown : $\tau_{dust}$ = 4 -- red : $\tau_{dust}$ = 1 -- orange : $\tau_{dust}$ = 0.4).
	       \textit{Top}: $\delta\theta$ = 3$^\circ$;
	       \textit{bottom-left}: $\delta\theta$ = 6$^\circ$;
	       \textit{bottom-right}: $\delta\theta$ = 10$^\circ$.}
     \label{Fig5.4}%
   \end{figure*}

In our first test, we fixed the absorbing equatorial optical depth to 40, which may be an upper limit for the medium opacity. A wider range of $\tau_{dust}$ is now investigated 
in Fig.~\ref{Fig5.4}, varying proportionally $\tau_{dust}$ from 40 to 0.4 (40, 4, 1, 0.4). The figure shows a series of intermediate cases between our results for an optically 
thick dust layer explored in the previous Sect.~\ref{Sect2.5.2} and the pure WHIM case (Sect.~\ref{Sect2.1}).

As seen from Fig.~\ref{Fig5.3}, at high optical depths of the dust, the scenarios of Fig.~\ref{Fig5.4} have similar polarimetric properties to the multi-component 
torus models investigated in \citet{Marin2012a}. The polarisation dichotomy is reproduced but the polarisation degree at type-2 inclinations is higher than 
observed $P$ levels \citep{Kay1994}. Decreasing the dust
optical depth ($\tau_{dust}$ = 4) diminishes the overall $P$ as photons can now escape from the dusty funnel by transmission through the dust wind. The polarisation dichotomy 
is preserved down to $\tau_{dust}$ approximatively equals unity, where the multiple regimes for parallel and perpendicular polarisation reappear. However, they leave 
a higher covering factor for the observed number density of type-2 AGN and the resulting polarisation degree at equatorial viewing angles is lower and thus closer 
to the observed range. Our modelling excludes optically thin dust environments ($\tau_{dust} <$ 1) as they automatically generate parallel photon polarisation
angles at equatorial inclinations. 

The impact of the model morphology is essentially visible at intermediate and equatorial viewing angles, where the observer's line-of-sight is crossing
the dust material. For type-1 objects, the polarisation degree is found to be parallel, with $P <$ 2~\%, slightly increasing with $i$. Around 45$^\circ$, 
$P$ remains stable for a few $i$, depending on the half-opening angle of the outflow. Such a polarisation plateau ($P \sim$ 3 -- 5~\%) associated with a perpendicular 
polarisation position angle is a strong observational prediction that could help to estimate the average $\delta\theta$ of BAL-quasars\footnote{Our predictions seem 
to be supported by the spectropolarimetric observations performed on 6 individual BAL QSO by \citet{Lamy2004a,Lamy2004b}, where $P$ is found to be 
of the order of a few percents, associated with a perpendicular polarisation position angle for 4 out of 6 objects.}. For higher inclinations, $P$ rises 
as backscattered photons from the WHIM funnel strongly contributes to the polarisation flux. When the internal structure becomes hidden by the cold, outer dust layer, $P$
starts to decrease until it reaches a minimal value along type-2 inclinations (below 10~\% at $\tau_{dust} \le$ 4).\\

We thus find that for $\theta = 45^\circ$, and $\delta\theta = 3^\circ$--$10^\circ$, and moderate optical depths of the dust with $1 < \tau_{dust} \le 4$, the double-wind 
model seems to rather well reproduce the observed continuum polarisation. It gives strong observational predictions on the geometrical opening angles of the 
outflow, as well as the amount of cold dust that can survive so close from the emitting source. However, it is important to investigate if the physical and 
geometrical changes to the model of \citet{Elvis2000} proposed are realistic and in agreement with number counts of different AGN types.


\section{Discussion}
\label{Sect3}

\subsection{Summarising our study of the outflow model}
\label{Sect3.1}

We tested to which extend the WHIM model of \citet{Elvis2000} can reproduce the observed polarisation dichotomy with respect to the polarisation position angle, 
as well as the expected level of polarisation percentage in type-1 and type-2 AGN. A pure Thomson scattering model with $\theta$ = 60$^\circ$ and 
$\delta\theta$ = 3$^\circ$ presents low $P$ levels ($<$ 1~\%) and cannot reproduce the perpendicular polarisation expected at equatorial inclinations. 
For low-luminosity objects, \citet{Elvis2000} predicts that the amount of polarisation along the outflow's direction, where a layer of dust prevents the
detection of BAL, would be below 0.5~\%. Our modelling agrees with this claim but adding dust to the outer parts of the conical outflows still does not produce 
the expected polarisation dichotomy. Finally, a double-layered wind also solely produces parallel polarisation, as well as too high polarisation percentages at 
type-2 viewing angles.

Exploring different sets of morphological and opacity parameters, we then found that small modifications help to match the modelling results with polarisation 
observations. A key point is to slightly lower the bending angle of the winds ($\theta_{60^\circ} \rightarrow \theta_{45^\circ}$) to allow for the presence 
of perpendicular polarisation at type-2 viewing directions. Perpendicular polarisation at intermediate and equatorial lines-of-sight emerges naturally when 
photons travel through geometrically thick regions with an optical depth equal or larger than unity.

For the pure WHIM model we found parametrisations that may correspond to the observed polarisation dichotomy between non-NAL/NAL (type-1) AGN and obscured BALQSO's 
or Seyfert-2 galaxies. However, the solid angle of the sky covered by Seyfert-2 galaxies may require additional dust obscuration that we also tested. 
Both, the observed polarisation dichotomy and polarisation degrees can be reproduced if we consider a dual wind model with $\theta$ = 45$^\circ$ and include 
equatorial scattering in the accretion flow. The half-opening angle $\delta\theta$ of the flow then normalises the polarisation degree as well as the maximum 
inclination at which parallel polarisation is still detected. The opacity of the outer (dusty) wind may also be adjusted to decrease the polarisation degree at
equatorial inclinations; the maximum polarised flux is then observed along the outflowing direction. To match with spectropolarimetric observations, the 
dust optical depth along the equator should be close to unity.

\subsection{Discriminating between different wind models}
\label{Sect3.2}

The question of discriminating between different wind models was raised by \citet{Elvis2000,Elvis2002a,Elvis2002b} discussing wind properties for the X-ray range. 
We here add a brief discussion about the UV/optical properties of the ionisation cones and narrow line regions seen in Seyfert-2 like galaxies. The first conical 
models were based upon the geometrical shape of the extended narrow line regions that are supposed to be the natural extension of the ionised outflows. 
The assumed morphology is an hourglass-shaped, bi-conical region with the emitting source situated at the cone base.

The resulting polarisation rises from face-on towards edge-on viewing angles and uniquely produces perpendicular polarisation angles, independently of the 
filling medium (either electrons or dust grains, \citealt{Goosmann2007}). Typically, $P$ can be as high as 20~\% for ionised or dusty winds. However, the wind 
structure of \citet{Elvis2000} presents a very different polarisation pattern: in contrast to polar outflows, the hollow winds can produce polarisation at parallel and 
perpendicular position angles. Compared with the model presented in Fig.~\ref{Fig5.3} and Fig.~\ref{Fig5.4}, the polarisation degree is small ($P <$~5~\%) at both polar
and equatorial viewing angles. The maximum polarisation degree is reached for lines-of-sight crossing the outflow, with 5 $< P <$ 25~\%, depending on the model's 
dust opacity and half-opening angle. The two-phase model succeeds to reproduce the polarisation degree expected from the polarised continuum of BAL 
quasars \citep{Cohen1995,Goodrich1995,Ogle1998}. 

The low-luminosity version of the wind with dusty extensions leads to $P<$0.5~\% as predicted for non-BAL objects \citep{Berriman1990}. Moreover, \citet{Elvis2000}'s 
two-phase model may produce a polarisation position angle that varies from 90$^\circ$ at pole-on views to 0$^\circ$ at equatorial views, which is in agreement with 
the observed polarisation dichotomy. Depending on the parametrisation, it is also possible to create $\psi$ = 0$^\circ$ photons at polar inclinations -- a phenomenon 
observed in some Seyfert-1 galaxies that were named \textit{polar scattering dominated} AGN by \citet{Smith2002}.

\subsection{Polarisation upper limits from Seyfert-atlases}
\label{Sect3.3}

Compared to the modelling results for AGN winds, the observed polarisation degrees are generally found to be significantly lower. In the atlas of Seyfert-2 galaxies 
presented by \citet{Kay1994}, a large polarisation degree was detected for Mrk~463E ($P$ = 10.29 $\pm$ 0.18~\% after starlight correction, $z$ = 0.050 
\citealt{Veron1987}), but finding such highly polarised type-2 AGN is quite uncommon. The majority of the observed Seyfert-2 galaxies show lower $P$ values, 
usually below 3~\% \citep{Miller1990,Kay1994}. Most Seyfert-1 galaxies have much weaker polarisation degrees, usually below 1~\% \citep{Smith2002,Smith2004}, except 
for exceptional, high-polarisation objects where $P$ can reach up to 5~\% (see Mrk~231 in \citealt{Goodrich1994} and Fairall~51 in \citealt{Smith2002}).

The diversity of detected polarisation degrees is difficult assess in the modelling. Simulations of simple and isolated \citep{Miller1990}, or complex and 
radiatively coupled \citep{Marin2012a,Marin2012b} reprocessing regions in AGN tend to create too large polarisation degrees at equatorial viewing angles and not 
enough polarisation at polar inclinations. The outflow model evaluated in this paper seems to be able to overcome these difficulties when taking into account a 
two-phase wind structure. High polarisation degrees at a perpendicular polarisation position angle can be reached if the inclination of the system is close to 
the bending angle of the quasar wind; low $P$ values are produced when the observer's line-of-sight is close to the equatorial plane. As the proposed set of 
outflow half-opening angles lies between 3$^\circ$ and 10$^\circ$, the ratio of high-$P$ to low-$P$ AGN remains small, which is coherent with the few 
high-$P$ Seyfert-2 galaxies detected so far. As for Seyfert-1 galaxies, the structure is able to produce parallel polarisation as high as 2~\% (Fig.~\ref{Fig5.3}), 
while preserving the observational evidence of higher $P$ values for equatorial viewing angles than at polar inclinations (Fig.~\ref{Fig5.3} and Fig.~\ref{Fig5.4}).

\subsection{The test case of the Seyfert galaxy NGC~5548}
\label{Sect3.4}

NGC~5548 is one of the best laboratories for testing AGN models \citep{Rokaki1993} and an important reference for \citet{Elvis2000}. How does the wind model hold 
against spectropolarimetric measurements of the object ? The Seyfert-1 galaxy ($z$ = 0.0174) shows a continuum polarisation degree of 0.67 $\pm$ 0.01~\% associated 
to a position angle of $\psi \sim$~45.6 $\pm$ 0.6 $^\circ$ \citep{Goodrich1994}\footnote{In comparison, \citet{Smith2002} found $P \sim$ 0.69 $\pm$ 0.01~\% and 
$\psi \sim$~33.2 $\pm$ 0.5 $^\circ$ for NGC~5548}. To compare our modelling predictions to the measured polarisation position angle, the observed $\psi$ needs 
to be related to the position angle of the radio axis of NGC~5548 (157$^\circ$). The angular difference of 111$^\circ$ is roughly perpendicular. 

According to \citet{Goodrich1994}, perpendicular polarisation indicates that polar electron scattering is the dominant scattering mechanism. However, 
\citet{Wu2001}'s estimation of the inclination towards NGC~5548 ($i$ = 47.3$^\circ$ $^{+7.6}_{-6.9}$) implies that the polarisation degree should be significantly 
higher than the observed value. Our model presented in Fig.~\ref{Fig5.1} can explain the polarisation degree and the polarisation position angle of NGC~5548 in 
the context of the wind model for a bending angle of 45$^\circ$. It strengthens the idea that AGN winds do not need to be fully filled by reprocessing matter; a 
hollow geometry may be then considered as more likely.

\subsection{The morphology of the outflow in IC~5063}
\label{Sect3.5}

Located in the southern hemisphere, IC~5063 ($z$ = 0.011, \citealt{Morganti2007}) is a Seyfert-galaxy that is famous to be the first AGN where a fast outflow of 
neutral hydrogen was detected \citep{Morganti1998}. Among other characteristics, its \textit{Hubble Space Telescope} image shows an ionised structure extending 
out to 15~kpc in a remarkable, X-shaped morphology \citep{Morganti2003}. However, spectroscopic and spectropolarimetric images of solid double cone models 
\citep{Wolf1999,Marin2012a} show that the scattering region re-emits uniformly to the irradiation source and the peculiar behaviour of IC~5063 cannot be visually 
explained using uniform or fragmented polar winds. The idea that AGN winds may be hollow came while observing the 3-dimensional velocity field of the line-emitting 
gas over 1~kpc in NGC~1365 \citep{Hjelm1996}. The flow pattern of such a wind then naturally creates an X-shaped signal in the imaging.

The WHIM model investigated in this paper (Fig.~\ref{Fig1.3} bottom) successfully reproduces\footnote{The \textit{Hubble Space Telescope} image presented by
\citet{Morganti2003} is based on total flux only, while our Fig.~\ref{Fig1.3} shows the polarised flux (total flux $\times$ polarisation degree). We checked that
the X-shape morphology is also reproduced in total flux, a straightforward conclusion as Thomson scattering is wavelength-independent.} the X-shaped morphology 
of IC~5063, as well as the bright flux spot in the AGN inner regions detected by \citet{Morganti2003}. It indicates that Thomson scattering is still a dominant
mechanism at large distances from the central source of IC~5063 and it tends to support the idea that quasar outflows are hollow structures. However, not all 
AGN where hollow winds are suggested show an X-shaped morphology (see for instance NGC~2992 in \citet{Veilleux2001} or NGC~1068 in \citet{Crenshaw2000}). 
A possible complication here may be that the amount of dust in the extended narrow line regions is high and thus prevents direct imaging of the hollow wind 
signatures.

\subsection{The tilted outflow of NGC~1068}
\label{Sect3.6}

NGC~1068 is considered to be one of the most archetypical examples of thermal Seyfert-2 galaxies. According to the simplest approach in AGN unification models 
\citep{Antonucci1993,Urry1995}, the central region of NGC~1068 is hidden by optically thick, circumnuclear matter usually designated as a dusty torus. It is then a 
direct consequence of the circumnuclear dust that the radiation of the central engine escapes non-isotropically along the polar regions. It forms the so-called 
ionised winds that farther out interact with the interstellar medium of the host galaxy and give rise to narrow emission lines \citep{Osterbrock1986}. It is a 
common hypothesis that as a consequence of the collimation effect by the circumnuclear dust the polar outflows sustain the same half-opening angle as the dusty torus.
This assumption is derived from the assumed axisymmetry of the unified model. However a recent study on NGC~1068 carried out by \citet{Raban2009} showed 
that the polar winds (represented by a bi-conical structure) are most likely inclined with respect to the obscuring torus axis. In the context of this work, several 
models of tilted outflows \citep{Das2006,Goosmann2011,Marin2012c} are being developed -- but the off-axis outflow may also be explained by the structure for quasars 
if the radiation pressure field is not isotropic. In this case, the bending of the flow becomes non-axisymmetric and the extended outflow would appear tilted 
with respect to the symmetry axis of the system.


\section{Conclusions and future work}
\label{Sect4}

Conducting detailed radiative transfer simulations that include polarisation, we show that the structure for quasars suggested by \citet{Elvis2000} can 
reproduce the observed optical polarisation properties of type-1 and type-2 AGN if the standard version of the model is slightly modified: the bending 
angle of the wind should be lowered to $\theta = 45^\circ$ and a dusty phase of moderate optical depth should be added to the wind.

Without the dusty phase, i.e. for a pure WHIM outflow, the polarisation dichotomy of Seyfert galaxies may still be reproduced for a certain, rather 
specific range of parameters but the predicted number density of type-2 AGN would turn out too low. \\

The modelling leads to some straightforward observational tests and future work:
\begin{enumerate}

\item The bending angle and covering factor of the extended winds that we derive from the polarisation modelling imply number densities for non-NAL and 
NAL AGN, BALQSO's, as well as type-1 versus type-2 AGN. Such number counts can be derived from multi-waveband data obtained in large sky surveys 
\citep[e.g.][]{Roseboom2013}, as the distribution of covering factors is correlated with the fraction of IR luminosity. It is thus possible to test 
if the observed distribution of different AGN types corresponds to their polarisation properties, and a preliminary investigation of the inclination-dependency 
of AGN spectropolarimetric observations is ongoing.

\item So far, our modelling only refers to UV/optical continuum radiation. Computations for emission and absorption lines that also consider different 
velocity fields inside the wind are currently under way. Furthermore, we intend to extend the modelling towards the X-ray regime and to compare the results 
to X-ray spectroscopy data.

\item The exact wind geometry varies most likely with the bolometric luminosity of the AGN. Therefore, it is useful to study the polarisation properties 
of well-observed, individual AGN with different average luminosities and to derive the resulting angles $\theta$ and $\delta\theta$ for the two-phase 
wind model. Possible trends of $\theta$ and $\delta\theta$ with the luminosity can thus be investigated.
\end{enumerate}

It is interesting to note that, from a purely morphological point of view, the two-phase version of the wind model by \citet{Elvis2000} is not in disagreement with the standard scenario of the unified scheme as analyzed in \citet{Marin2012a}. Nonetheless, it is a major difference between the two interpretations to assume that the obscuring material originates in a wind instead of being part of the accretion inflow.


\section*{Acknowledgments}

The authors are grateful to the referee Makoto Kishimoto for his useful and constructive comments on the manuscript. This research was funded by the French 
grant ANR-11-JS56-013-01 of the project POLIOPTIX. The authors a grateful to Delphine Porquet for clarifying comments on the manuscript.

\label{lastpage}

\end{document}